\documentclass[12pt]{article}
\usepackage{amsfonts,amsthm,amsmath,amssymb,times}
\usepackage{fullpage,graphicx,multicol}
\usepackage{epic,eepic}

\usepackage[small]{caption}

\newcommand{\paperone}{\cite{Lin+al07}}

\newcommand{\includegraphicsbw}[2]{\includegraphics[#2]{#1.eps}}
\renewcommand{\includegraphicsbw}[2]{\includegraphics[#2]{#1_bw.eps}}

\newcounter{kevinslistcounter}%
\renewenvironment{itemize}{
\begin{list}{(\alph{kevinslistcounter})}
{\usecounter{kevinslistcounter}
\setlength{\parsep}{0pt}
\setlength{\labelwidth}{24pt}
\setlength{\itemsep}{0pt}
\setlength{\topsep}{\parsep}}}{\end{list}}

\newcounter{kevinslistcountertoo}%
\renewenvironment{enumerate}{
\begin{list}{\arabic{kevinslistcountertoo}.}
{\usecounter{kevinslistcountertoo}
\setlength{\parsep}{0pt}
\setlength{\labelwidth}{24pt}
\setlength{\itemsep}{6pt}
\setlength{\topsep}{\parsep}}}{\end{list}}

\newtheorem{proposition}{Proposition}[section]
\newtheorem{definition}{Definition}[section]

\newtheorem{theorem}{Theorem}
\newtheorem{lemma}{Lemma}[section]

\newcommand{\tht}{\theta}

\newcommand{\eps}{\varepsilon}
\newcommand{\aff}{a_{\rm ff}}

\newcommand{\bec}{\begin{center}}
\newcommand{\eec}{\end{center}}
\newcommand{\bei}{\begin{itemize}}
\newcommand{\eei}{\end{itemize}}
\newcommand{\beq}{\begin{equation}}
\newcommand{\eeq}{\end{equation}}
\newcommand{\beqr}{\begin{eqnarray}}
\newcommand{\eeqr}{\end{eqnarray}}
\newcommand{\beqrn}{\begin{eqnarray*}}
\newcommand{\eeqrn}{\end{eqnarray*}}
\newcommand{\brr}{\begin{array}}
\newcommand{\err}{\end{array}}
\newcommand{\bef}{\begin{figure}}
\newcommand{\eef}{\end{figure}}

\newcommand{\om}{\omega}

\newcommand{\lmax}{\lambda_{\rm max}}

\date{August 22, 2007}

\title{Reliability of Coupled Oscillators II:\\ Larger Networks}

\author{Kevin K. Lin$^{1}$, Eric Shea-Brown$^{1,2}$, and Lai-Sang Young$^{1}$ \\ \\
$^1$ Courant Institute of Mathematical Sciences,  \\
$^2$ Center for Neural Science \\
New York University, New York, NY 10012, U.S.A. }

\begin{document}

\maketitle

\begin{abstract}
We study the \textit{reliability} of phase oscillator networks in
response to fluctuating inputs. Reliability means that an input elicits
essentially identical responses upon repeated presentations, regardless
of the network's initial condition.  In this paper, we extend previous results
on two-cell networks to larger systems.  The first issue that arises is
chaos in the absence of inputs, which we demonstrate and interpret in
terms of reliability.  We give a mathematical analysis of networks that can be
decomposed into modules connected by  an
acyclic graph.  For this class of networks, we show how
to localize the source of unreliability, and address questions concerning
downstream propagation of unreliability once it is produced.
\end{abstract}
%\note{second half changed}

%%%%%%%%%%%%%%%%%%%%%%%%%
\begin{small}
\tableofcontents
\end{small}

\section*{Introduction}
\addcontentsline{toc}{section}{Introduction}

This is a sequel to {\it Reliability of Coupled Oscillators
  I}~\paperone, but can be read independently (if the reader is willing
to take a few facts for granted).  Both papers are about the reliability
of coupled oscillator networks in response to fluctuating
inputs. Reliability here refers to whether a system produces identical
responses when it is repeatedly presented with the same stimulus.  The
context in this paper is that of heterogeneous networks of coupled phase
oscillators. The stimuli, or input signals, are assumed to be white
noise; they are received by some components of the network and relayed
to others with or without feedback.
%\note{rephrased to minimize repeats}

There has been substantial progress in understanding the reliability of
single oscillators
~\cite{piko01,Nak+al05,Zho+03,Ter+03,Rit+03,Pak+01,Pak+03,Gol+05,Gol+06}.
Paper I contains a detailed study of reliability properties for the
simplest possible network, that of two coupled cells \paperone. We
showed that this small network displays both reliable and unreliable
behaviors in broad parameter ranges. Here, we consider larger networks
building on some of the insight gained from Paper I.
%\note{move Japanese ref to end of intro}

Our first observation concerns intrinsic network chaos, meaning the
network in question can be chaotic even in the absence of any inputs.
This phenomenon occurs only in systems consisting of 3 or more
oscillators, and the chaos can be sustained in time and ubiquitous in
the phase space.  Its implications are discussed in Sects.~2 and 6.

Our main technical analysis is in Sect.~3. It is first carried out for
acyclic networks, that is, networks whose coupling connections have no
cycles, and then extended to a larger class obtained by replacing the
nodes in acyclic networks by  {\it modules} (which are themselves
networks).  It is at this latter class that much of the discussion
in the rest of the paper is aimed.
%\note{deleted "small"}

In a larger network, there is a need to understand the problem locally
as well as globally: when unreliability occurs, it is useful to know
where it is created and which parts of the network are affected.
Therefore, instead of solely using a single index, namely the largest
Lyapunov exponent $\lmax$, as indicator of reliability for the entire
system, we have sought to localize the source of unreliability, and to
devise ways to assess the reliability of outputs registered at any
given site.  Among our findings is that once produced, unreliability
propagates; no site downstream can then be completely reliable.

In studying networks that decompose into smaller modules, there is
a reassembly issue, namely how the information gained
from studying the modules separately can be leveraged in
the analysis of the full network. We will present some evidence
to show that such a strategy may be feasible.
%\note{added paragraph}

Numerous examples are given to illustrate the ideas proposed.  In
addition to the 2-oscillator system studied in \paperone, we present a
few other examples of small networks capable of both reliable and
unreliable behavior, finding some rather curious (and unexplained) facts
along the way.

The mathematical analysis in Sect.~3 is rigorous, as are the facts
reviewed in Sects.~2.2 and 4.1. The rest of the investigation is
primarily numerical, supported whenever possible by facts from dynamical
systems theory.  In the course of the paper we have highlighted a number of observations, based in
part on empirical data, that are directly relevant for the broader
theoretical questions at hand.

There is a vast literature on networks of oscillators; it is impossible
to give general citations without serious omissions. We have limited our
citations to a sample of papers and books with settings closer to ours
or which treat specifically the topic of reliability (see below).  We
mention in particular the recent preprint \cite{Ter+07}, which discusses
the reliability of large, sparsely coupled networks in a way that
complements ours.
%\note{new paragraph}

%%%%%%%%%%%%%%%%%%%%%%%%%%%%%%%%%%%%%%

\section{Model and measure of reliability}
\label{model formulation}

\subsection{Coupled phase oscillator systems}

\begin{figure}
\begin{center}
\includegraphics[scale=0.5]{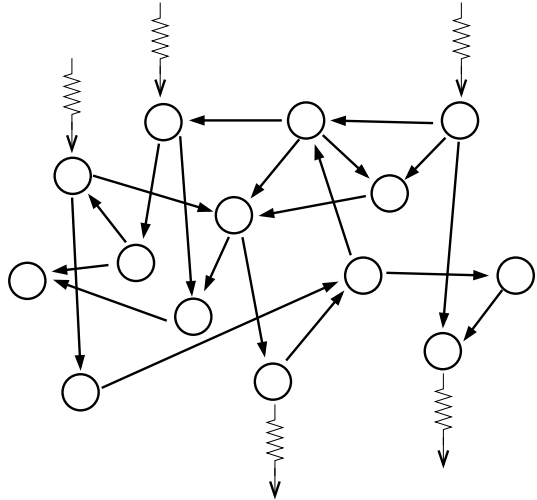}
\end{center}
\caption{Schematic of a forced oscillator network.}
\label{f.general_network}
\end{figure}

We consider in this paper networks of coupled phase oscillators in
the presence of external stimuli.  A schematic picture of such a setup
is shown in Figure~\ref{f.general_network}. The networks considered
are of the type that arise in many settings; see {\it e.g.},
\cite{piko01,SS00,Krev,HopIzhik,geomtime,excit,BHMsiro,Erm91rev,EKmult}.
%\note{changed -- to separate from technical description}

The {\it unforced} systems, {\it i.e.}, the systems without external
stimuli, are described by
\begin{equation}
  \dot\theta_i \ = \ \omega_i \ + z(\theta_i)\ \sum_{j \neq
    i} a_{ji}\ g(\theta_j)\ , \qquad i = 1, \cdots, N\ .
\label{noinput}
\end{equation}
In this paper, $N$ can be any number $\ge 1$; in particular, it can be
arbitrarily large.
%\note{added sentence to signal how to think of $N$}
The state of oscillator $i$ is described by an angular variable
$\theta_i \in S^1 \equiv {\mathbb R}/{\mathbb Z}, i=1, \cdots, N$. Its
intrinsic frequency is given by a constant $\omega_i$.  We allow these
frequencies to vary from oscillator to oscillator.
%, but assume for
%definiteness that they are all within about $10\%$ of $1$.
The second term on the right represents the coupling: $a_{ji} \in
{\mathbb R}$, $g$ is a ``bump function'' vanishing outside of a small
interval $(-b,b)$; on $(-b,b)$, it is smooth, $\geq 0$, and satisfies
$\int_{-b}^b{g(\theta)\ d\theta} = 1$; $z(\theta)$ is a function which
in this paper is taken to be $z(\theta)
=\frac{1}{2\pi}(1-\cos(2\pi\theta))$.  The meaning of this coupling
term is as follows: We say the $j$th oscillator ``spikes'' when
$\theta_j(t)=0$. Around the time that an oscillator spikes, it emits a
pulse which modifies some of the other
oscillators~\cite{peskin,StrMir90,HopHer95,NumPer85}. The sign and
magnitude of $a_{ji}$ describe how oscillator $j$ affects oscillator
$i$: $a_{ji}>0$ (resp. $a_{ji}<0$) means oscillator $j$ excites (resp.
inhibits) oscillator $i$ when it spikes, and $a_{ji}=0$ means
oscillator $i$ is not directly affected. In this paper, $b$ is taken to
be about $\frac{1}{20}$, and the $a_{ji}$ are taken to be ${\mathcal
O}(1)$. Finally, the function $z(\theta)$, often called the phase
response curve~\cite{Winf74,G75,erme96,Krev}, measures the variable
sensitivity of an oscillator to coupling and stimulus inputs.

The {\it stimulus-driven} systems are of the form
\begin{align}
 d\theta_i &= \omega_i\ dt \ + z(\theta_i)\ \Big( \sum_{j \neq
   i} a_{ji}\ g(\theta_j)\ dt + I_i(t)\Big)\ , \nonumber\\
    I_i(t) & = \eps_i\ dW_i(t)\ , \ \qquad \qquad \qquad \qquad i = 1, \cdots, N\ .
\label{input}
\end{align}
Here, $I_i(t)$ is the external stimulus received by oscillator $i$; it
is taken to be white noise with amplitude $\epsilon_i$.  For
simplicity, we assume that for $i \ne j$, $dW_i(t)$ and $dW_j(t)$ are
either independent, or they are one and the same, {\it i.e.}, the same
input is received by both oscillators although the amplitudes may
differ. As discussed earlier, we are primarily interested in situations
where inputs are received by only a subset of the oscillators, with
$\epsilon_i\approx 0$ for the rest.  Likewise, we are sometimes
interested in the response registered at only a subset of the oscillators rather
than the whole network.  See Fig.~\ref{f.general_network}.

For more on the motivation and interpretation of this model,
see Sect. 1.2 of ~\cite{Lin+al07}.

%%%%%%%%%%%%%%%%%%%%%%%%%%
\subsection{Lyapunov exponents as indicator of reliability}

{\it Reliability} in this paper refers to the following: Suppose the
stimuli in Eq. (\ref{input}) described by the terms $\varepsilon_i I_i(t)$
are turned on at time $t = 0$. Will the response of the system
depend on the initial state of the system when the stimuli are presented?
That is to say, let ${\bf \Theta}(t)$ be the solution of Eq. (\ref{input}).
To what degree does ${\bf \Theta}(t)$ depend on ${\bf \Theta}(0)$
after some initial period of adjustment?
If the dependence on ${\bf \Theta}(0)$ is negligible after a transient,
we think of the system is {\it reliable}; and if different
${\bf \Theta}(0)$ lead to persistently different responses at time $t$
for all or most $t$,
we say the system is {\it unreliable}. Needless to say, the lengths of
the transient and the degree to which the responses differ are important
in practical situations, but these issues will not be considered here.

The measure of reliability used in a good part of this paper is $\lmax$,
the largest Lyapunov exponent of the system. We briefly recall here the
meaning of this number and its relation to reliability as discussed
above, referring the reader to~\cite{RDS,Bax92} for background material
on stochastic flows and to Sect. 2 of~\cite{Lin+al07} for a more
detailed discussion of how this material is used in the study of
reliability.
%\note{reinserted references}

We view Eq. (\ref{input}) as a stochastic differential equation (SDE)
on ${\mathbb T}^N \equiv S^1 \times \cdots S^1$ with phase variables
$(\theta_1, \cdots, \theta_N)$. A well known theorem gives the
existence of a {\it stochastic flow of diffeomorphisms}. That is
to say, for almost every realization $\omega$ of white noise, there
exist flow maps of the form $F_{t_1,t_2;\omega}$: here $t_1<t_2$ are
any two points in time, and $F_{t_1,t_2;\omega}({\bf \Theta}(t_1,
\omega))={\bf
  \Theta}(t_2, \omega)$ where ${\bf \Theta}(t,\omega)$ is the solution
of (\ref{input}) corresponding to $\omega$.  By the Lyapunov exponents
of Eq. (\ref{input}), we refer to the Lyapunov exponents of
$F_{0,t;\omega}$ as $t$ increases.  These numbers do not depend on
$\omega$; and they do not depend on ${\bf \Theta}(0)$ if
Eq. (\ref{input}) is ergodic (which it often is). The largest Lyapunov
exponent is denoted by $\lmax$.

In ~\cite{Lin+al07}, {\it cf.}~\cite{Rit+03,Gol+06,Pak+03}, we used of $\lmax$ as a measure of
reliability.  Specifically, $\lmax<0$ is equated with reliability,
$\lmax>0$ represents unreliable behavior, and $\lmax=0$ is
inconclusive.\footnote{We note that there are other ways to measure
  unreliability which seem to us to be equally reasonable.  For example,
  one can take the sum of all {\em positive} Lyapunov exponents. }
  These criteria will continue to be used in this
paper.

To understand why $\lmax$ is a reasonable indicator of reliability,
recall the idea of {\it sample measures}: if we let $\mu$ denote the
stationary measure of (\ref{input}), {\em i.e.} the stationary solution
to the Fokker-Planck equation, the sample measures are then the
conditional measures of $\mu$ given the past history of the stimuli.
More precisely, if we think of $\omega$ as defined for all $t \in
(-\infty,\infty)$ and not just for $t>0$, then $\mu_\omega$ describes
what one sees at $t=0$ given that the system has experienced the input
defined by $\omega$ for all $t <0$.  Sample measures thus give an
approximation of what one sees in response to an input presented on a
time interval $[-t_0, 0)$ for large enough $t_0>0$.  If we let
  $\sigma_t$ denote the time-shift of the Brownian path $\omega$, then
  sample measures evolve in time according to
  $(F_{0,t;\omega})_*\mu_\omega = \mu_{\sigma_t\omega}$.  As explained
  in ~\cite{Lin+al07}, if stimuli corresponding to a realization
  $\omega$ are turned on at $t=0$, then $\mu_{\sigma_t \omega}$ gives a
  good approximation of the responses of the system at sufficiently large times $t$ for an
  ensemble of initial conditions sampled from a smooth density.  In
  particular, if $\mu_{\sigma_t \omega}$ collapses down to a single
  point as $t$ increases, then we conclude that initial conditions have
  little effect after transients, {\em i.e.} the system is reliable.  If
  $\mu_{\sigma_t \omega}$ is spread out nontrivially in the phase space,
  and this condition persists with $t$, then we conclude that different
  initial conditions have led to different responses at time $t$, which
  is exactly what it means for the system to be unreliable.

The relation between sample measures, which represent a collection of
the system's possible responses, and $\lmax$, its top Lyapunov
exponent,
 is summarized in the following general
result from dynamical systems theory.

\begin{theorem} Consider a SDE with an ergodic stationary measure $\mu$.
\begin{itemize}
\item[(1)] \cite{LeJ85}  If
$\lambda_{\max}<0$, then with probability 1, $\mu_\omega$ is supported
on a finite set.
\item[(2)] \cite{Led+88}
If $\mu$ has a density and
$\lambda_{\max}>0$, then with probability 1, $\mu_\omega$ is a
random SRB measure.
\end{itemize}
\end{theorem}

Case (1) is often referred to as the system having {\it random sinks}.
When $\lmax<0$, in the vast majority of situations $\mu_\omega$ is
supported on a single point; we therefore equate it with reliability as in, e.g.,~\cite{Rit+03,Pak+01}.
In Case (2), the system is said to have a {\it random strange
attractor}.  For deterministic systems, SRB measures are well studied
objects with a distinctive geometry: they have densities on (typically)
lower dimensional surfaces that wind around the phase space in a
complicated way; see~{\cite{Eck+85}}.  Here we have a random ({\em
i.e.} $\omega$-dependent) version of the same picture.  Thus the
resulting geometry is a readily recognizable signature of
unreliability.

Finally, we remark that $\lmax$ is a {\it global} indicator of
unreliability, meaning a positive exponent will tell us only that the
network possesses some unreliable degrees of freedom: it does not
specify where this unreliability is produced, whether its effects are
localized, and so on.  Different measurements are needed to
answer these more refined questions. They are discussed in
Sects. 3--6.

%%%%%%%%%%%%%%%%%%%%%%%%%%%%%%%%
%%%%%%%%%%%%%%%%%%%%%%%%%%%%%%%%

\section{Intrinsic network chaos}
\label{intrinsic chaos}

%Once we leave the class of acyclic networks, unreliability can occur
%(as we have seen in ~\cite{Lin+al07}).

\subsection{Chaotic behavior in undriven networks}

This section discusses an issue that did not arise in ~\cite{Lin+al07}.
Our criterion for unreliability, $\lmax>0$, is generally equated with
{\it chaos}. In the 2-oscillator network studied in~\cite{Lin+al07},
this chaotic behavior is triggered by the input: without this forcing
term, the system cannot be chaotic as it is a flow on a
two-dimensional surface. The situation is quite different for larger
networks.

\bigskip
    \noindent {\bf Observation 1:}  {\it Networks of 3 or more
pulse-coupled phase oscillators can be chaotic in the
absence of any external input.}

\bigskip We begin by explaining what we mean by ``chaotic.'' It is well
known that systems of 3 or more coupled oscillators without external
input can have complicated orbits, such as homoclinic orbits or
horseshoes (see, {\it e.g.}, \cite{Pop+05,NumPer85}, and also
\cite{gh,Erm91rev} for general references).
%\note{deleted ref to $2$-D; sentence too long}
These special orbits by themselves do not necessarily constitute a
seriously chaotic environment, however, for they are not always easily
detectable, and when horseshoes coexist with sinks, the chaotic behavior
seen is transient for most choices of initial conditions.  We claim here
that networks with 3 or more oscillators can support a stronger form of
chaos, namely that of {\it strange attractors}.  There is no formal
definition of the term ``strange attractors'' in rigorous dynamical
systems theory. An often used characterization is the existence of SRB
measures or physical measures, which describe the long-time distribution
of orbits starting from positive Lebesgue measure sets of initial
conditions {\cite{Eck+85, lsy-srb, lsy-jsp}}.  At a minimum, one would insist that
positive Lyapunov exponents be observed for almost all -- or at least on
a large positive Lebesgue measure set -- of initial conditions.  (In the
case of horseshoes coexisting with sinks, for example, Lyapunov
exponents for most initial conditions are negative after a transient.)
In numerical simulations, one has to settle for positive Lyapunov
exponents that are sustained over time for as many randomly chosen
initial conditions as possible. This will be our operational definition
of {\it intrinsic network chaos}.

One of the simplest configurations that support intrinsic network chaos
is
\begin{equation}
\includegraphics[scale=0.5]{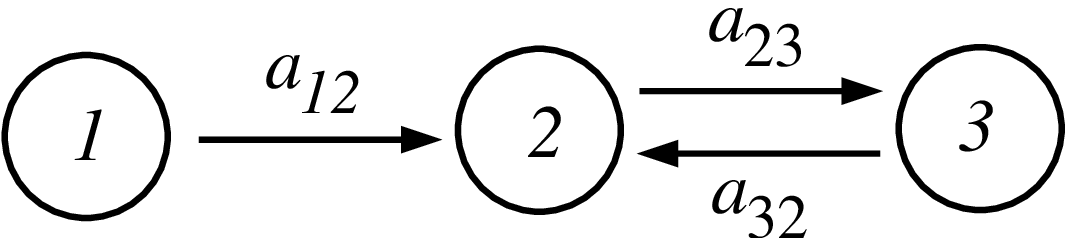}
\label{intrinsic chaos example}
\end{equation}
The setup above is similar to that studied in \paperone, except that
the (2,3)-subsystem is kicked {\it periodically} by pulses from
oscillator 1 rather than driven by an external stimulus in the form of
white noise. Ignoring the presence of oscillator 1 for the moment, we
saw in \paperone, Sect.~4.2, that when $a_{23}$ and $a_{32}$ are
roughly comparable and away from $0$, the phase space geometry of
oscillators 2 and 3 is favorable for {\em shear-induced chaos}, a
geometric mechanism for producing chaotic behavior. Indeed, a rigorous
theory of this mechanism gives the existence of strange attractors with
SRB measures when such a system is periodically kicked --- provided the
kicks are suitably directed and time intervals between kicks are
sufficiently large~\cite{WY2,WY3}. Even though the kick intervals
(of $\approx 1$)
here are too short for rigorous results to be applicable, numerical
simulations show that there are definitively positive exponents
sustained over long periods of time when $a_{12}$ is large enough.

As an example, consider system (\ref{intrinsic chaos example}) with $\omega_1,
\omega_2, \omega_3 = 0.93, 1, 1.1$, and $a_{23}, a_{32} =1, 1.45$. We
find that $\lmax$ increases from about $0.025$ to $0.12$ for
$a_{12}\in(1,1.5)$.\footnote{These Lyapunov exponents are computed
  starting from $\sim 10$ initial conditions and have an absolute error
  of roughly 0.005.  } Notice that the system
parameters above are chosen so that the (2,3)-subsystem is unreliable,
that is, so that $\lmax>0$ when oscillator 2 receives a white noise
stimulus: see \paperone, Fig.~12.  It is, after
all, the same mechanism, namely shear-induced chaos, that produces the
chaos in both situations. See \cite{ly07}, Study 4, for similar
findings.

While there are certainly large classes of networks that do not exhibit
intrinsic network chaos ({\em e.g.} see Sect.~\ref{acyclic}), from the
example above one would expect that such chaos is commonplace among
larger networks.  Moreover, for intrinsically chaotic networks, one would expect $\lmax$
  to remain positive when relatively weak stimuli are added to various
  nodes in the network. We have, in fact, come across a number of
  examples in which $\lmax$ remains positive for larger stimulus
  amplitudes.  This raises the following issue of interpretation: In an
  intrinsically chaotic network that also displays $\lmax>0$ in the
  presence of inputs, it is unclear how to attribute the source of
  unreliability, since the effects of inputs and intrinsic chaos are
  largely inseparable.  We adopt here the view that whatever the cause,
  such a network is unreliable.

%%%%%%%%%%%%%%%%%%%%%%%
%%%%%%%%%%%%%%%%%%%%%%%
\subsection{Suppression of network chaos by inputs: a case study}

An interesting question is whether or not networks with intrinsic chaos
of the form described above, {\em i.e.} with $\lmax>0$ for large sets
of initial conditions in the absence of inputs, necessarily respond unreliably to external stimuli.  %\note{changed sentence}
We find that the answer is ``no'': we have come across a
number of instances where weakly chaotic networks are stabilized by
sufficiently strong inputs.  An example is in the following 3-cell
ring:
\begin{equation}
\includegraphics[scale=0.45]{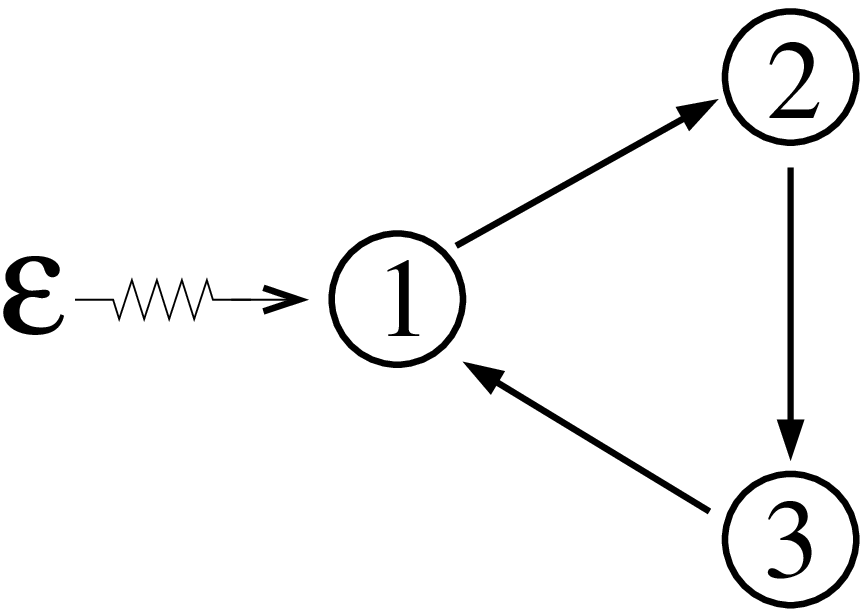}
\label{f.threecell}
\end{equation}

\noindent We find small regions of parameter space near $a_{12}=1,
a_{23}=2, a_{31}=0.6$ with intrinsic network chaos, as evidenced by
positive values of $\lmax$: see Fig.~\ref{f.3-ring
intrinsic chaos suppression}(a). %\note{deleted note: info in Fig suffices}
As the reader will notice, $\lmax$
oscillates quite wildly in this region, with $\lmax>0$ and $\lmax =0$
occurring at parameter values in close proximity. This is
characteristic of a large class of {\it deterministic} dynamical systems as we
now explain.

\begin{figure}[t]
% src: ~/hopf/rely1.data/study03/2007-07-24.d/paperfig.m
% src: ~/hopf/rely1.data/study03/2007-07-19.d/paper-figs.ss
\begin{center}
\resizebox{!}{2.8in}{\includegraphicsbw{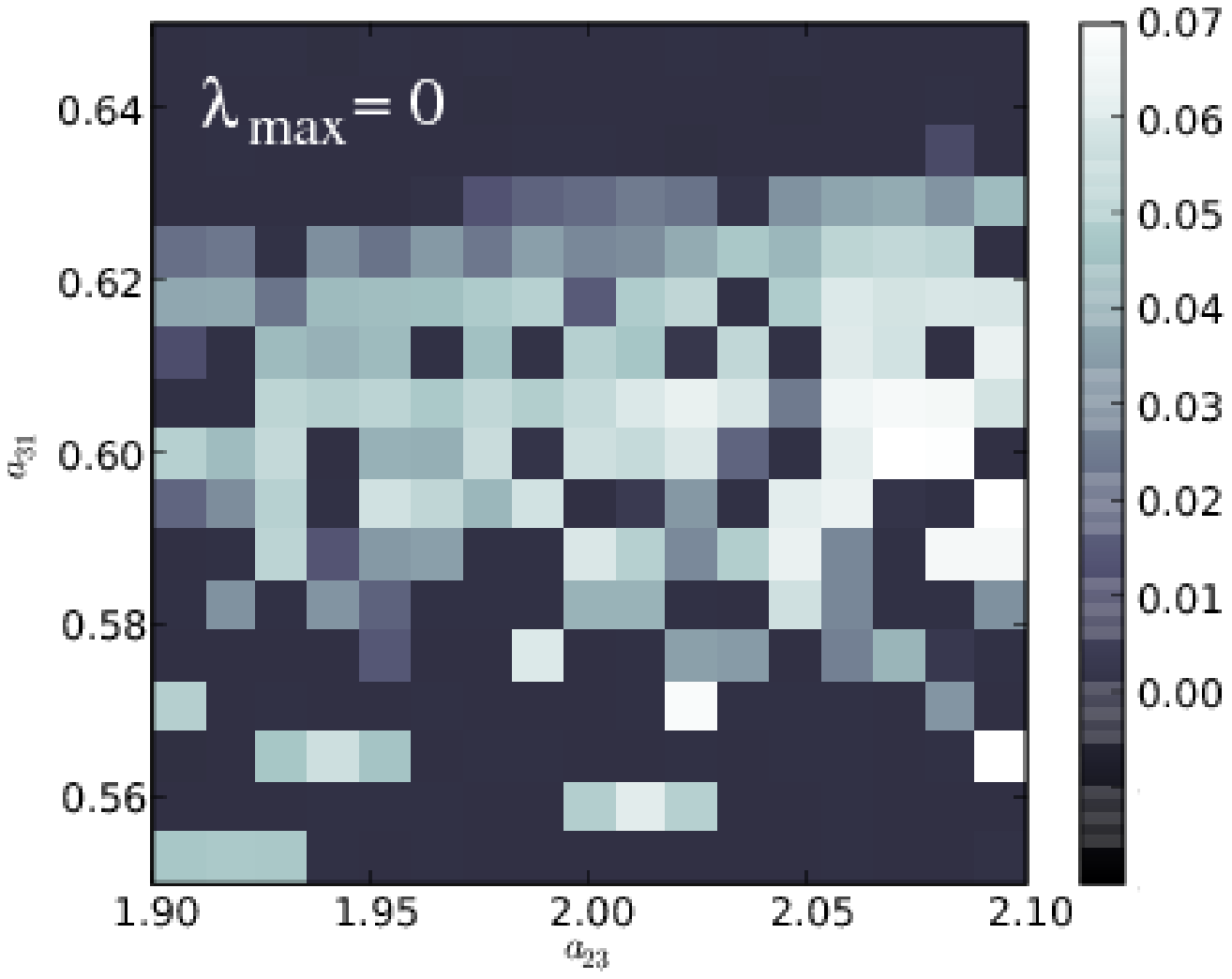}{bb=0 0 434 332}}%
\includegraphics[bb=0in 0in 2.8in 2.3in,scale=1.1]{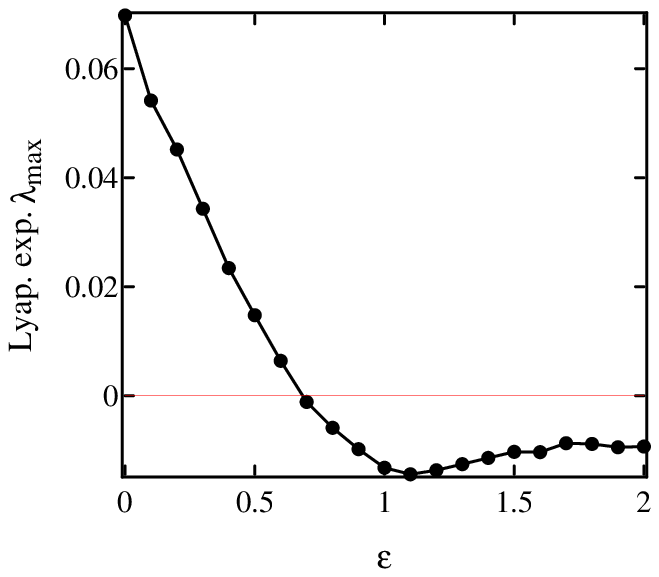}
\end{center}
\caption{Suppression of intrinsic chaos in Eq.~(\ref{n-ring}).  {\bf
    Left:} the Lyapunov exponent $\lmax$ with $\eps=0$ is shown as a
  function of $a_{23}$ and $a_{31}$ ($a_{12}$ is fixed at 1).  Each
  exponent is computed using 4 initial conditions.  For the vast
  majority ($\geq 95\%$) of the points computed, the error in the
  computed exponent is $\lesssim 0.005$.  {\bf Right:} we show the
  response to $\eps>0$ for $a_{23}=2.075$ and $a_{31}=0.6025$.  In all
  plots, we use $\omega_1=1$, $\omega_2=0.95$, and $\omega_3=1.1$. }

\label{f.3-ring intrinsic chaos suppression}
\end{figure}

The phase space of system (\ref{f.threecell}) is the $3$-torus.  Taking
the section $\{\theta_1=0\}$ of the $\eps=0$ flow,
 we obtain a return map $\Psi$ from the $2$-torus to
itself.  There are well known examples of maps of the $2$-torus that
are uniformly hyperbolic; a case in point is the ``cat map''.  These
maps are robustly hyperbolic; it is impossible to destroy their
positive Lyapunov exponents by reasonable-size perturbations.  Our
section map $\Psi$, however, cannot be of this type for topological
reasons: it is not hard to see that $\Psi$ can be deformed continuously to the
identity map by suitably tuning system parameters.  Generally speaking,
for $2$-D maps that can be deformed to the identity, the only way to
form strange attractors is by ``folding''.  Such systems are at best
{\it nonuniformly hyperbolic}; their observed Lyapunov exponents are
the results of competing tendencies: there is expansion in the phase
space, which necessarily leads to contraction elsewhere or in different
directions, and the expanding and contracting directions are not
separated as in the cat map.  As a result of this competition,
sustained, observable chaos, {\it
  i.e.}, strange attractors, live side by side in parameter space with
transient chaos, {\it i.e.}, systems in which horseshoes and sinks
coexist.  The simplest family of examples exhibiting such a phenomenon
is the logistic family~\cite{Lyu03,Jak81,Gra+97}. This phenomenon also
occurs in shear-induced chaos; see {\it e.g.}, \cite{ly07}.
For an exposition on nonuniformly hyperbolic systems, see
{\it e.g.}, \cite{lsy-srb}. The mixed behavior alluded to
above is discussed in, {\it e.g.} \cite{lsy-jsp}.

Thus even without explicit knowledge of the map $\Psi$, one would expect
to see a pattern of behavior similar to that in Fig.~\ref{f.3-ring
  intrinsic chaos suppression}(a) when positive exponents are present.
Here, $\lmax=0$ for the flow corresponds to $\Psi$ having a Lyapunov
exponent $\le 0$, including the parameters with transient chaos.

We now consider the effects
of adding a stimulus, that is, taking $\eps>0$ for
system~\eqref{f.threecell}.  Informally, one may think of the stimulus
as sampling nearby parameters, averaging (in a loose sense) the
different tendencies, with chaos suppression made possible by the
presence of the mixed behavior.  A sample of our numerical results is
shown in Fig.~\ref{f.3-ring intrinsic chaos suppression}(b). At
$\eps=0$, $\lmax >0$, indicating network chaos. As expected, $\lmax$
remains positive for small $\eps$.  But, as $\eps$ increases, $\lmax$
steadily decreases and eventually becomes mildly negative,
demonstrating that sufficiently strong forcing can suppress intrinsic
network chaos and make a network more reliable.  A possible mechanistic
explanation is that when $\eps$ is large enough, oscillator 1 is
affected more by the stimulus than by oscillator 3 due to the
relatively weak coupling $a_{31}$. As in the case of a single
oscillator, the stimulus has a stabilizing effect on oscillator 1,
making it reliable.

Our understanding from the discussion above may be summarized
as follows:

 \bigskip
 \noindent {\bf Observation 2:} {\it Some networks with weak intrinsic chaos will
   respond reliably to moderately strong stimuli.}

%%%%%%%%%%%%%%%%%%%%%%%%%%%%%%%%
%\section{Architecture Conducive to Reliability}
\section{Acyclic networks and modular decompositions}
\label{acyclic}

Since the possibilities in completely general networks are vast, we
think a good place to start may be the class of {\it acyclic} networks.
These are networks in which there is a well defined direction of
information flow.  We will show that acyclic networks are not
intrinsically chaotic, and they are never unreliable. Building on the
analysis developed for acyclic networks, we identify a broader class
that may be accessible to analysis, namely networks that admit a
decomposition into modules with acyclic inter-module connections.

For simplicity, we assume
throughout that the stimuli are independent;
it is trivial to modify the results of this section to accomodate the
situation when some of them are identical to each other.

\subsection{Skew-product representation of acyclic networks}
\label{sect2.1}

We first describe the connection graph that corresponds to an oscillator
network. Let each node of this graph, $i \in \{1, \cdots, N\}$,
correspond to an oscillator.  Assign a directed edge from node $i$ to
node $j$, $i \ne j$, if oscillator $i$ provides input to oscillator $j$,
{\em i.e.} if $a_{ij} \neq 0$; in this case, we write $i \to j$.  (For
simplicity, we do not allow nodes to connect to themselves.)  A {\it
  cycle} in such a directed graph is a sequence of nodes and edges $i_1
\to i_2 \to \cdots i_k \to i_1$ for some $k > 1$.

\begin{definition}
We say an oscillator network is {\it acyclic} if its connection graph
has no cycles.
\end{definition}

Given any pair of oscillators in an acyclic network, either they are
``unrelated'', or one is ``upstream'' from the other. We say oscillator
$i$ is ``upstream'' from oscillator $j$ if there is a sequence of nodes
and edges that goes from $i$ to $j$. The absence of cycles is
precisely what makes this ``upstream''--``downstream'' notion well
defined. We say oscillators $i$ and $j$ are ``unrelated'' if there is no chain
that goes from $i$ to $j$ or {\it vice versa}. Unrelated oscillators
do not necessarily behave independently: they may,
for example, receive input from the same source. Overall,
the structures of acyclic networks can still be quite complex,
with tree-like branching and recombinations possible.

Our first task is to find a systematic way to treat the dynamics within
these networks.

\begin{lemma} In an acyclic graph, one can define for each node $j$
a number $m(j)$ representing its maximum number of ancestors, meaning

(i) there is a chain of the form $i_1 \to i_2 \to \cdots \to i_{m(j)}
\to j$, and

(ii) there is no chain of the form $i_1 \to i_2 \to \cdots \to
i_{m(j)+1} \to j$.
\end{lemma}

\noindent The proof is simple: chains cannot be arbitrarily long
without a node repeating, and such repeats are impossible in acyclic
graphs.

\medskip
Using the notation of Sect. 1.1, we now discuss the
dynamical structure of acyclic oscillator networks.
First, let
$\varphi_t$ denote the flow on ${\mathbb T}^N$ in the absence of inputs,
{\em i.e.}, with $\eps_i \equiv 0$.  We say $\varphi_t$ {\it factors
  into a hierarchy of skew-products with 1-dimensional fibers} if
after a permutation of the names of the $N$ oscillators, we have
the following: For each $k=1, \cdots, N$, there is a vector field $X^{(k)}$ on
${\mathbb  T}^k$ such that if $\varphi_t^{(k)}$ is the flow generated by
$X^{(k)}$, then (i) $\varphi_t^{(k)}$ describes the dynamics of the network
defined by the first $k$ oscillators and the relations among them, and
(ii) $\varphi^{(k+1)}_t$ is a {\it skew-product}
over $\varphi^{(k)}_t$, that is, the vector field $X^{(k+1)}$ on
${\mathbb T}^{k+1}$ has the form
\begin{equation}
X^{(k+1)} (\tht_1, \cdots, \tht_{k+1}) = (X^{(k)}(\tht_1, \cdots,
\tht_k), Y_{(\tht_1, \cdots, \tht_k)}(\tht_{k+1})) \label{skew}
\end{equation}
where $\{Y_{(\tht_1, \cdots, \tht_k)}\}$ is a family of vector fields on
$S^1$ parametrized by $(\tht_1, \cdots, \tht_k) \in {\mathbb T}^k$.
In particular, $\varphi^{(N)}_t=\varphi_t$. In
the system defined by (\ref{skew}), we refer to $\varphi_t^{(k)}$ on
${\mathbb T}^{k}$ as the flow on the {\it base}, and each copy of $S^1$
over ${\mathbb T}^{k}$ as a {\it fiber}.

\begin{proposition} The flow of every acyclic network of $N$ oscillators
with no inputs can be represented by a hierarchy of skew-products with
1-dimensional fibers.
\label{lemma: hierarchy skew products}
\end{proposition}

\noindent {\bf Proof:} Let ${\cal N}_m$ be the set of oscillators $j$
with $m(j)=m$.  Assign an ordering to the oscillators so that all the
oscillators in ${\cal N}_0$ come first, followed by those in ${\cal
N}_1$, then ${\cal N}_2$, and so on. The order within each ${\cal N}_m$
is immaterial. Let us denote this ordering by $i_1 < i_2 < \cdots i_N$.
A skew-product is built inductively as follows: For $k \geq 1$,
consider the subnetwork consisting of oscillators $i_1, \cdots, i_k$.
None of the oscillators $i_j, j \leq k$, receives input from any of the
oscillators $i_\ell, \ell>k$. (To see this, note that $i_\ell \to i_ j$
implies $m(i_j) > m(i_\ell)$ by definition, but this is impossible
under the ordering we have chosen.) Therefore, the dynamics of
oscillators $i_1, \cdots, i_k$ as a subsystem of the entire network may be
described by a vector field of the form $X^{(k)}(\tht_{i_1}, \cdots,
\tht_{i_k})$, obtained by projecting $X^{(N)}$ onto these $k$ coordinates.
Starting with $k=1$, the skew-products in the lemma are constructed
inductively for increasing $k$.
%\note{previous version had wrong indices}
\hfill $\square$

\bigskip

Next we generalize to acyclic networks with stimuli.  Such networks can
also be represented by a directed graph of the type described above,
except that some of the nodes correspond to stimuli and others to
oscillators. If $i$ is a stimulus and $j$ an oscillator, then $i \to j$
if oscillator $j$ receives information directly from stimulus $i$. No
arrow can terminate at a stimulus, so that $m(i)=0$ if $i$ is a
stimulus. Clearly, a network driven by stimuli is acyclic if and only
if the corresponding network without stimuli is acyclic.

Proceeding to skew-product representations, consider first the case of
a single oscillator driven by a single stimulus. Let $\Omega$ denote
the set of all Brownian paths defined on $[0,\infty)$, and let
$\sigma_t: \Omega \to \Omega$ be the time shift. Then the dynamics of
the stochastic flow discussed in Sect. 2.1 can be represented as the
skew-product on $\Omega \times S^1$ with
$$
\Phi_t : (\omega, x) \ \mapsto \  (\sigma_t(\omega),
F_{0,t;\omega}(x))\ .
$$
Similarly, a network of $N$ oscillators driven by $q$ independent
stimuli can be represented as a skew-product with base $\Omega^q$
(equipped with product measure) and
fibers ${\mathbb T}^N$.

\begin{proposition} The dynamics of an acyclic network driven by $q$
  stimuli can be represented by a hierarchy of skew-products over
  $\Omega^q$ with 1-dimensional fibers.
\label{p.heirarchydriven}
\end{proposition}

\noindent The proof is identical to that of Proposition~\ref{lemma: hierarchy skew
  products}, except that when enumerating the nodes of the graph, we
first list all of the stimuli (in any order) before listing the
oscillators.

%%%%%%%%%%%%%%%%%%%%%%%%%%%%%%%%

\subsection{Lyapunov exponents of acyclic networks}
\label{sect2.2}

Consider a network of $N$ oscillators driven by $q$ independent
stimuli. We first review  some notation.
Let $\omega \in \Omega^q$ denote a $q$-tuple of Brownian paths, and let
$F_{0,t;\omega}$ denote the corresponding stochastic flow on ${\mathbb
T}^N$.  For a fixed $\omega \in \Omega^q, x \in {\mathbb T}^N$ and
tangent vector $v$ at $x$, define the {\it
  Lyapunov exponent}
\begin{equation}
\lambda_\omega(x,v) \ = \ \lim_{t \to \infty} \ \frac{1}{t} \log
|DF_{0,t;\omega}(x)v|
\label{def lyapunov exponent}
\end{equation}
if this limit exists. If $\mu$ is a stationary measure of the stochastic
flow, then for a.e. $\omega$ and $\mu$-a.e. $x$, $\lambda_\omega(x,v)$
is well defined for all $v$. The following is the main result of this
section:

\begin{theorem}
  Consider a network of $N$ oscillators driven by $q$ independent
  stimuli, and let $\mu$ be a stationary measure for the stochastic
  flow. Assume

    \begin{itemize}
    \item the network is acyclic, and

    \item $\mu$ has a density on ${\mathbb T}^N$.
    \end{itemize}

\noindent Then $\lambda_\omega(x,v) \leq 0$ for a.e. $\omega \in
\Omega^q$ and $\mu$-a.e. $x$.
\end{theorem}

\noindent
One way to guarantee that condition (b) is satisfied is to set $\eps_i$
to a very small but strictly positive value if oscillator $i$ is not
originally thought of as receiving a stimulus, so that $\eps_i>0$ for
all $i$.  Such tiny values of $\eps_i$ have minimal effect on the
network dynamics. Condition (b) may also be satisfied in many cases
where some $\eps_i=0$ if suitable hypoellipticity conditions are
satisfied, but we do not pursue this here
{\cite{hypoelliptic}}.

Before proceeding to a proof, it is useful to recall the following
facts about Lyapunov exponents. For a.e. $\omega$ and $\mu$-a.e. $x$,
there is an increasing sequence of subspaces $\{0\}=V_0 \subset V_1
\subset \cdots \subset V_r = {\mathbb R}^N$ and numbers $\lambda_1 <
\cdots < \lambda_r$ such that $\lambda_\omega(x,v) = \lambda_i$ for
every $v \in V_i \setminus V_{i-1}$. The subspaces depend on $\omega$
and $x$, but the exponents $\lambda_j$ are constant a.e. if the flow is
ergodic.
% , and
%\begin{equation}
%\lim_{t \to \infty} \ \frac{1}{t} \log |\det(D\varphi_{\omega,t}(x))|
%  \ = \ \sum_{i=1}^{r(x)} \lambda_i m_i \label{det}
%\end{equation}
%where
We call a collection of $N$ vectors $\{v_j\}$ a {\it Lyapunov basis} if
exactly $\dim(V_i) - \dim(V_{i-1})$ of these vectors are in $V_i
\setminus V_{i-1}$. If $\{v_j\}$ is a Lyapunov basis, then for any $u, v
\in \{v_j\}$, $u\neq v$,
\begin{equation}
    \lim_{t \to \infty} \ \frac{1}{t}  \log |\sin \angle(DF_{0,t;\omega}(x)u,
DF_{0,t;\omega}(x)v)| = 0 ,
\label{e.sineq}
\end{equation}
that is, angles between vectors in a Lyapunov basis do not decrease
exponentially fast; see {\it e.g.},
%(otherwise (\ref{det}) would be violated)
~\cite{lsy-srb} for a more detailed exposition.

\bigskip
\noindent {\bf Proof:} Since the network is acyclic, it factors into a
hierarchy of skew-products.  The $k$th of these is a stochastic flow
$F^{(k)}_{0,t;\omega}$ on ${\mathbb T}^k$. It describes the (driven)
dynamics of the first $k$ oscillators assuming they have been reordered
%\note{clarified ordering}
so that oscillator $i$ is upstream from or unrelated to oscillator $j$ for $i<j$.
Let $\mu^{(k)}$ denote the projection of $\mu$ onto ${\mathbb T}^k$. Then
$\mu^{(k)}$ is an stationary measure for $F^{(k)}_{0,t;\omega}$, and it
has a density since $\mu$ has a density. We will show inductively
in $k$ that the conclusion of
Theorem 2 holds for $F^{(k)}_{0,t;\omega}$.

%\note{rephrased}
For $k=1$, $\lambda_\omega(x,v) \leq 0$ for a.e. $\omega$ and
$\mu^{(1)}$-a.e. $x$. This is a consequence of
Jensen's Inequality; see {\it e.g.}, \paperone, Sect. 2.2, for more
detail.

Assume we have shown the conclusion of Theorem 2 up to $k-1$,
and view $F^{(k)}_{0,t;\omega}$ as
a skew-product over $\Omega^q \times {\mathbb T}^{k-1}$ with
$S^1$-fibers. Choose a vector $v_k$ in the direction of the
$S^1$-fiber. Note that this direction is invariant under the
variational flow $DF^{(k)}_{0,t;\omega}$ due to the skew-product
structure. Starting with $v_k$, we complete a Lyapunov basis $\{v_1,
\cdots, v_k\}$ at all typical points. Due to the invariance of the
direction of $v_k$, we may once more use Jensen's Inequality to show
that $\lambda_\omega(x,v_k) \le 0$ for a.e. $x$ and $\omega$.  We next
consider $v_i$ with $i < k$.
First, define the projection $\pi: {\mathbb T}^k \to {\mathbb
T}^{k-1}$, and note that
$$| DF^{(k)}_{0,t;\omega}(x)v_i | = \frac{ | \pi
  (DF^{(k)}_{0,t;\omega}(x)v_i)| } { |\sin \angle(v_k,
  DF^{(k)}_{0,t;\omega}(x)v_i)|} .$$ Due to Eq.~\eqref{e.sineq}, we have
$\lambda_\omega(x,v_i) = \lim_{t \rightarrow \infty} \frac{1}{t} \log |
\pi (DF^{(k)}_{0,t;\omega}(x)v_i) |$.  But the skew-product structure
yields $\pi (DF^{(k)}_{0,t;\omega}(x)v_i) = DF^{(k-1)}_{0,t;\omega}(\pi
x) (\pi v_i)$, so by our induction hypothesis, $\lambda_\omega(x,v_i)
\leq 0$. \hfill $\square$

\bigskip \noindent {\bf Remarks on reliability of acyclic networks:} Our
conclusion of $\lmax \leq 0$ falls short of reliability, which
corresponds to $\lmax <0$.  When there are free-rotating oscillators in
a network, {\em i.e.} oscillators that are not driven by either a
stimulus or another oscillator, then clearly $\lmax=0$. When no
free-rotating oscillators are present, typically one would expect $\lmax
<0$. We do not have a rigorous proof, but this intuition is supported by
numerical simulations.

\bigskip
Arguments similar to those in the proof of Theorem 2 (but in the absence
of stimuli and without the use of an invariant measure) give the following:

\begin{proposition}
  Acyclic networks are never intrinsically chaotic, in the sense that at
  Lebesgue-$a.e. x \in {\mathbb T}^N$, all Lyapunov exponents (with
  $\limsup$ instead of {\rm limit} in Eq.~(\ref{def lyapunov
    exponent})) are $\le 0$.
\end{proposition}

%%%%%%%%%%%%%%%%%%%%%%%%%%%%

%\subsection{Quotient systems and reliability}
\subsection{Modular decompositions and quotient systems}
\label{sect2.3}

We next describe how the reliability of more general networks may be
analyzed by decomposition into subunits.
Consider a graph with nodes $\{1, \cdots, N\}$ as
in the beginning of Sect.~\ref{sect2.1}, and let $\sim$ be an
equivalence relation on the set $\{1, \cdots, N\}$. The quotient graph
defined by $\sim$ has as its nodes the equivalence classes $[i]$ of
$\sim$, and we write $[i] \to [j]$ if there exists $i' \in [i]$ and $j'
\in [j]$ such that $i' \to j'$. The following is a straightforward
generalization of Proposition \ref{p.heirarchydriven}:

\begin{proposition} In a network of oscillators driven by $q$
  independent stimuli, if an equivalence relation leads to an acyclic
  quotient graph, then the dynamics of the network can be represented by
  a hierarchy of skew-products over $\Omega^q$, with the dimensions of
  the fibers equal to the sizes of the corresponding equivalence
  classes.
  \label{p.quotientskew}
\end{proposition}

We pause to discuss in more detail the structure in
Prop~\ref{p.quotientskew}, as it is important in the rest of this
paper. An equivalence relation on the nodes of a network describes a
decomposition of the network into smaller subunits called {\em
  modules}. Introducing directed edges between modules as above, we
obtain what we call a {\em quotient network} in this
paper.  Assume this quotient network is acyclic, and let $M_1, M_2,
\cdots, M_p$ be the names of the modules ordered in such a way that
$M_i$ is upstream from or unrelated to $M_j$ for all $j>i$. Let $k_j$ be
the number of nodes in module $M_j$. For $s_i=k_1 + k_2 + \cdots + k_i$,
let $F^{(s_i)}_{0,t,\omega}$ denote, as before, the stochastic flow
describing the dynamics within the union of the first $i$ modules; we do
not consider $F^{(s)}_{0,t,\omega}$ when $s \ne s_i$ for some $i$.  The
dynamics of the entire network can then be built up layer by layer as
follows: We begin with the stochastic flow $F^{(k_1)}_{0,t,\omega}$.
Then proceed to $F^{(k_1+k_2)}_{0,t,\omega}$, which we view as a
skew-product over $F^{(k_1)}_{0,t,\omega}$. This is followed by
$F^{(k_1+k_2+k_3)}_{0,t,\omega}$, which we view as a skew-product over
$F^{(k_1+k_2)}_{0,t,\omega}$, and so on.

Let $\lambda^{(1)}_1, \cdots, \lambda^{(1)}_{k_1}$ denote the Lyapunov
exponents of $F^{(k_1)}_{0,t,\omega}$. Clearly, these are the Lyapunov
exponents of a network that consists solely of module $M_1$ and the
stimuli that feed into it. We wish, however, to view $M_1$ as part of
the larger network. If $\lambda^{(1)}_{\max} \equiv \max_j
\lambda^{(1)}_j >0$, we say {\it unreliability is produced within
$M_1$}. An interesting question is the effect of this unreliability on
sites downstream. Leaving this to Sect. 4, we continue with the present
discussion: For $i>1$, let $\lambda^{(i)}_1, \cdots,
\lambda^{(i)}_{k_i}$ denote the {\it fiber Lyapunov exponents} in the
skew-product representation of $F^{(s_i)}_{0,t,\omega}$ over
$F^{(s_{i-1})}_{0,t,\omega}$, and let $\lambda^{(i)}_{\max} = \max_j
\lambda^{(i)}_j$. Then $\lambda^{(i)}_{\max} >0$ has the interpretation
that {\it unreliability is produced within module $M_i$  as it operates in the larger network}. It
is important not to confuse this with the Lyapunov exponents of module
$M_i$ in isolation, an issue we will follow up in Sect. 6.

Analogous interpretations for the zero-input systems are obvious:
for $i>1$, $\lambda^{(i)}_{\max}>0$ at $\eps=0$ means
there is intrinsic network
chaos within the module $M_i$ as a subsystem of the larger system,
and so on.

The proof of
the following result is virtually identical to that of Theorem 2:

\begin{proposition} Suppose for a driven network there is an equivalence
  relation leading to an acyclic quotient graph. Then, with respect to any ergodic
  stationary measure $\mu$, the numbers $\lambda^{(i)}_j, 1 \le i \le p, 1 \le j
  \le k_i$, are precisely the Lyapunov
  exponents of the network.
 \label{c.decomp}
\end{proposition}

\noindent Proposition~\ref{c.decomp} says in particular that if, in
each of the $p$ skew-products in the hierarchy, the fiber Lyapunov
exponents are $\le 0$, {\it i.e.}, if no unreliability is produced
within any of the modules, then $\lmax$ for the entire network is $\le
0$. Conversely, if unreliability is produced within any one of the
modules as it operates within this network, then $\lmax>0$ for the
entire network. On the practical level, the skew product structure
(which implies that $DF_{0,t;\omega}$ is block-lower-triangular) and
the Proposition together give a more efficient way to numerically
compute Lyapunov exponents of networks with acyclic
quotients.

\bigskip
\noindent {\bf Important remarks.}   %\note{Rephrased - to keep message simple}
{\it In the rest of this paper, we
  will often take the view that the network in question is equipped with
  a modular decomposition connected by an acyclic graph.} Such a
decomposition always exists for any network, but it can be trivial ({\it
  e.g.}, when the entire network is a single module).  If the
  decomposition is nontrivial and $\lmax>0$ for the network,
  Proposition~\ref{c.decomp} enables us to localize the {\it source}
  of the unreliability, {\it i.e.}, to determine in which module
  unreliability is produced via their fiber Lyapunov exponents.
In particular, modules that are themselves
acyclic cannot produce unreliability.

  % or not very useful,
%such as when we are not better equipped to treat the individual modules.
%While it is not technically an assumption, we are naturally more
%interested in cases where the modules are tractable, hence relatively
%small. This means that the network is primarily feedforward, with
%feedback connections permitted ``locally''.\note{``with feedback
%  connections...''} The issues and ideas in Sects. 4 and 6 are
%particularly relevant for networks in this category.

\bigskip
As noted earlier, the idea of ``upstream''--``downstream'' for
acyclic networks extends to modules connected by acyclic graphs,
so that it makes sense to speak of a module as being
downstream from another module, or a site as being downstream
from another site, meaning the modules in which they reside are
so related.

%%%%%%%%%%%%%%%%%%%%%%%%%%%%%%%%%%%
%%%%%%%%%%%%%%%%%%%%%%%%%%%%%%%%%

\section{Propagation of unreliability}
\label{Propagation of unreliability}

In this section, we address the following basic question: {\it Under
what conditions will unreliability generated in one part of a network
propagate downstream? } In Sect.~\ref{site reliability}, we discuss how
to measure unreliability at specific network sites, and in
Sect.~\ref{downstream}, we address the question posed.

%\subsection{Site distribution and entropy}
\subsection{Measuring reliability at individual network sites}
\label{site reliability}

Often, it is the response of a network measured at specific oscillators
(or {\it sites}) that is of relevance, rather than the response of the
network as a whole. While $\lmax>0$ tells us that there is unreliability
somewhere in the system, it does not tell us which oscillators are affected.
%\note{less abrupt intro to $\mu_{\omega, i}$}
To capture the idea of reliability at individual sites, recall that
the reliability of the entire system is reflected in the sample measures
$\mu_\omega$ (see Sect. 1.2). By the same reasoning, the reliability at
site $i$ is reflected in the marginals of $\mu_\omega$ in the variable
$\theta_i$; we denote this marginal distribution by $\mu_{\omega, i}$.
We say a network is {\it reliable at site $i$} if $\mu_{\omega, i}$ is
concentrated at a single point; the more uniformly distributed these projected
sample measures are, the greater the {\it site unreliability}.
These ideas are easily generalized to groups of more than one site,
but we will treat only single site reliability.

The following are three standard ways in which the distribution of
$\mu_{\omega, i}$ can be described:

\medskip
\noindent {\it A. Site entropy.} \ For each $i$ and $\omega$, we let
$H(i,\omega)$ denote the entropy of the distribution $\mu_{\omega, i}$,
{\it i.e.}, if $\rho_{\omega,i}$ is the density of $\mu_{\omega, i}$ with
respect to Lebesgue measure on $S^1$, then   %\note{more accurate math}
\begin{displaymath}
H(i, \omega) = -\int_{S^1}{\log\rho_{\omega,i}\ d\rho_{\omega,i}}\ ,
\end{displaymath}
and we set $H(i,\omega)=-\infty$ if $\mu_{\omega, i}$ is singular.
The {\it site entropy} $H(i)$ is defined to be the expected value of this
random variable, {\it i.e.}, $H(i)= \int H(i, \omega) P(d\omega)$. In
practice, $H(i)$ is computed as
$\lim_{T\to\infty}\frac{1}{T}\int_0^T{H(i,\sigma_t(\omega))\ dt}$ via
the Ergodic Theorem. This number can range from $-\infty$ to $0$,
%\note{slight rephrasing}
with ${H}(i)=0$ corresponding to
uniform distribution and ${H}(i)=-\infty$ corresponding to the
distribution being singular with respect to Lebesgue measure.  A
drawback of site entropy is that it does not distinguish among singular
distributions.

\medskip
\noindent {\it B. Information dimension.} We define
$$
D(i,\omega) = \lim_{k \to \infty} D_k(i,\omega) \qquad {\rm where}
\qquad D_k(i,\omega) =
\frac{- \sum_{j=1}^k p_j \log p_j}{\log k}\ .
$$
In the quantity on the right, $S^1$ is divided into $k$ equal
intervals, and $p_j$ is the probability with respect to
$\mu_{\omega,i}$ of finding the phase of oscillator $i$ in the $j^{th}$
subinterval. The relevant quantity is then $D(i)=\int D(i,\omega)
P(d\omega)$.  This takes values on $[0,1]$, with $D(i)=1$ corresponding
to any distribution having a density.   Information dimension does not, for example,
distinguish between the uniform distribution on $S^1$ and a
distribution supported uniformly on a tiny subinterval of $S^1$.

\medskip
\noindent {\it C. Cumulative distribution functions (CDFs).} The most
direct way to assess site distributions is to compute the CDF of
$\mu_{\omega, i}$. This is numerically very simple to do, and is
especially effective in establishing whether the measure is
concentrated at a single point. A drawback of using CDFs is that
it is not a number. Moreover, to be certain that one is seeing
``typical'' CDFs, one needs to compute them for $\mu_{\omega,i}$
for many values of $\omega$, as  CDFs cannot be averaged.

\medskip Above, we discussed various ways to assess
$\mu_{\omega,i}$ from a purely theoretical standpoint.
 Returning to the situation at hand, recall from Theorem
1 that when $\lmax>0$, {\em i.e.} when the system is unreliable,
$\mu_\omega$ is a random SRB measure. These measures have densities on
complicated families of smooth manifolds. In particular, they have
dimensions $>1$; {\it e.g.} in the case of 2 oscillators, the dimension
of the random SRB measures is $1+\alpha$ for some $\alpha \in (0,1)$ where
$\alpha$ describes the dimension of the fractal part.
 A well known result in analysis \cite{mattila} tells us
that when measures of dimensions $>1$ are projected onto 1-dimensional
subspaces, the projected measures have a density for projections along
{\it almost all} directions. Now not all projections are ``good'' in
this sense, and we cannot be certain that the projection onto any
particular site is ``good'' (projections onto sites upstream from where
unreliability is produced are obviously not). Still, {\it if} the
projection to site $i$ is good, then $\lmax>0$ would imply that
$\mu_{\omega, i}$ has a density.  This being the case, it is more
important to be able to compare different
distributions with densities than to distinguish different singular
distributions.
%distinguish those with densities that to compare from those without.
 We therefore favor site entropy over information
dimension as a measure of site reliability.

In this paper, we will use a combination of site entropy and CDFs: If
the computed values of $H(i)$ appear bounded below,
%\note{deleted $t \to \infty$: no context}
then it is safe to conclude
site unreliability, and the closer $H(i)$ is to $0$, the more
unreliable. If, on the other hand, the computed values of $H(i)$ appear
unbounded, then CDFs are used to confirm site reliability.

%%%%%%%%%%%%%%%%%%%%%%%%%
\subsection{Sites downstream from unreliable modules}
\label{downstream}

\begin{figure}
\includegraphics[bb=0in 0in 2.2in 2in]{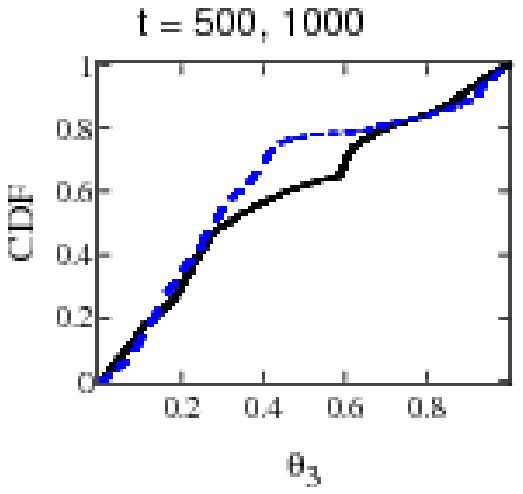}%
\includegraphics[bb=0in 0in 2.2in 2in]{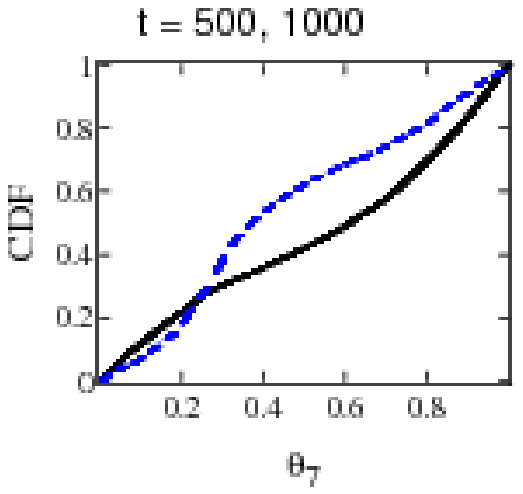}%
\includegraphics[bb=0in 0in 2.2in 2in]{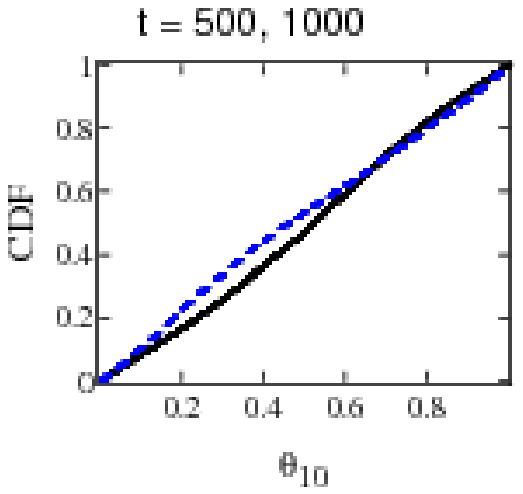}\\

\caption{Site distributions for the system~(\ref{propagation example})
  with $N=10$.  Shown are the site-$i$ CDFs for $i=3, 7, 10$, at times
  $t=500$ (solid) and $t=1000$ (dash).  The parameters are as follows:
  for the (1,2) subsystem, we set $\omega_1=1$, $\omega_2=1.05$,
  $a_{12}=1$, $a_{21}=1.18$, and $\eps=1$ (so that $\lmax\approx 0.13$).
  For $i \geq 3$, we draw $\omega_i$ from $[0.9,1.1]$ and $a_{i,i+1}$
  from $\pm[0.6,1.2]$.}
\label{cdfs for propagation of unreliability}
\end{figure}

We now return to the question of propagation of unreliability to
sites downstream from an unreliable module.  The simplest network with
which to investigate this is the $N$-chain system
\begin{equation}
\includegraphics[scale=0.6]{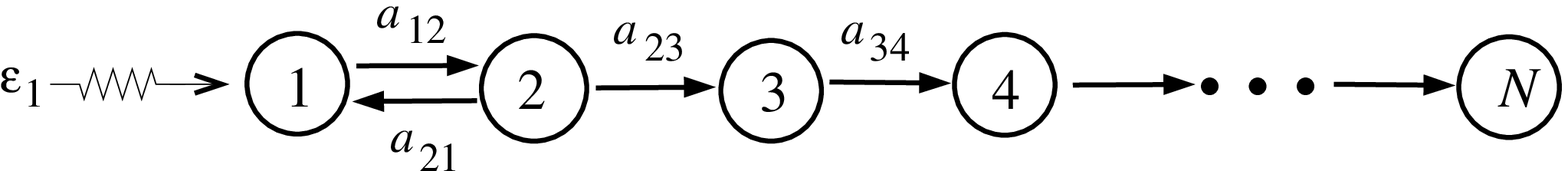}
\label{propagation example}
\end{equation}
Suppose we choose $a_{12}$, $a_{21}$, and $\eps_1$ so that the
(1,2)-subsystem is unreliable. Notice that no unreliability is produced
elsewhere. The question is: will this unreliability be observable
downstream, at sites $i=3, 4, \cdots, N$, or does it somehow
``dissipate?''  We run the following numerical experiment: we set
$\omega_1=1$, $\omega_2=1.05$, $a_{12}=1$, $a_{21}=1.05$, and $\eps_1=1$,
so that the Lyapunov exponent $\lmax$ of the (1,2)-subsystem is
$\approx 0.1$. For $i \geq 3$, we draw randomly (and fix) $\omega_i$
from $[0.9,1.1]$ and $a_{i,i+1}$ from $\pm[0.6,1.2]$. Computing the
site entropies\footnote{We compute site entropies by simulating the
response
  of $10,000$ initial conditions to the same realization of the
  stimulus, then applying the ``binless'' estimator of Kozachenko and
  Leonenko {\cite{kozachenko}} (see also {\cite{victor}}). }
  ${H}(i)$ for sites $i=3, 7, 10$, we find:
\begin{center}
\begin{tabular}{c|c|c|c}
Site & $i=3$ & $i=7$ & $i=10$\\\hline
Entropy ${H}(i)$ & -0.4 & -0.2 & -0.01\\
\end{tabular}
\end{center}
To help interpret these numbers, recall that identifying $S^1$ with $[0,1]$,
the entropy of the uniform distribution on an interval $[a,b]$ is
$\log(b-a)$, and $\log \frac{1}{2} \approx -0.7$.
The data above thus indicate that the site distributions are fairly
uniform, and in fact seem to become more uniform for sites farther
downstream. Fig.~\ref{cdfs for propagation of unreliability} shows the
corresponding CDFs for some sites at representative times; CDFs at
other sites are qualitatively similar.  The graphs clearly show that
unreliability propagates.

We give two explanations for why one should expect this result of
propagating site unreliability.  The first is a plausibility argument
along the lines of the projection argument in Sect.~\ref{site
reliability}: it is possible -- but highly unlikely -- that the SRB
measure $\mu_\om$ would project to point masses in any of the 8
directions corresponding to the 8 sites downstream. A second, perhaps
more intuitive explanation, is as follows.  Consider site 3 in the
system~(\ref{propagation example}). Fix a realization of the stimulus,
and let $(\theta_1, \theta_2, \theta_3)$ denote the 3 phase
coordinates.  For each choice of $(\theta_1(0), \theta_2(0))$, the
third oscillator receives a sequence of coupling impulses from the
(1,2)-subsystem.  Since the (1,2)-subsystem is unreliable, different
choices of $\theta_1(0)$ and $\theta_2(0)$ will produce different
sequences of coupling impulses to oscillator 3, in turn leading to
different values of $\theta_3(t)$.  This is synonymous with site
unreliability for oscillator 3.

\medskip What happens if an oscillator receives inputs from more than
one source with competing effects? The simplest situation is an
oscillator driven by both an unreliable module upstream and an input
stimulus, as depicted in Diagram (\ref{interference example 1}).  Note
the presence of competing
terms: As we have just seen, the unreliable module leads to
unreliability at site 3.  However, the stimulus $\eps_3$ has a
stabilizing effect on that oscillator.
\begin{equation}
\label{interference example 1}
\includegraphics[bb=0 0 362 75,scale=0.5]{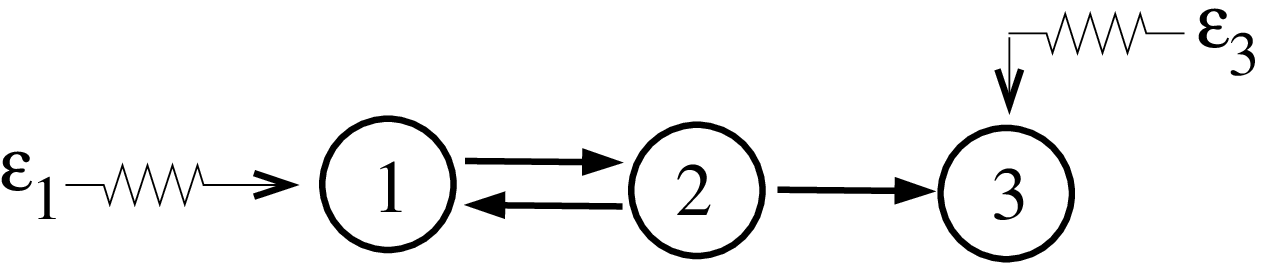}
\end{equation}
In the results tabulated below, the parameters used are $\omega_1=1$,
$\omega_2=1.05$, $a_{12}=1$, $a_{21}=1.15$, and $\eps_1=1$ (so that the
(1,2)-subsystem is again unreliable), $a_{23}=0.5$ and $\omega_3=0.93$.
The site entropy ${H}(3)$ is computed for various values of $\eps_3$.
We find:
\begin{center}
\begin{tabular}{c|c|c|c}
Stimulus amplitude &  $\eps_3=0.2$ & $\eps_3=1$ & $\eps_3=2$\\\hline
Site entropy ${H}(3)$ & -0.4 & -1.2 & -2\\
\end{tabular}
\end{center}
These numbers tell us that at $\eps_3=0.2$, the distribution at site 3
is fairly uniform, and that this distribution becomes more concentrated
as $\eps_3$ increases.  The CDFs (not shown) confirm this.  When
$\eps_3=2$, for example, about 80\%
of the distribution $\mu_{\omega,3}$ is concentrated on an interval of
length $\leq 1/5$ roughly 70\% of the time.  These data show that the
source of reliability, {\it i.e.}, the stimulus into oscillator 3,
attenuates the propagation of unreliability; we call this phenomenon
{\em interference}. Moreover, we find that when $\eps_3$ is increased
further, the support of a large fraction of $\mu_{\omega,3}$ shrinks to
smaller and smaller intervals, decreasing the entropy and reflecting a
greater tendency to form random sinks in oscillator 3. However,
simulations also show that oscillator 3 does not become fully site
reliable even at fairly strong forcing.  This is also expected:
intuitive arguments similar to those above suggest that once created,
site unreliability cannot be completely destroyed at downstream sites.
To summarize:

\bigskip
\noindent {\bf Observation 3:} {\it  Unreliability, once generated, propagates
to all sites downstream.}

\bigskip
We finish by clarifying the relation between the material
in Sect. 3.3 and this section:

\medskip
\noindent {\bf Propagation {\it versus} production of
  unreliability:}
The topic of Sect. 4 is whether or not unreliability created upstream
propagates, {\em i.e.}, whether it can be observed downstream.  This
concept complements an idea introduced in Sect. 3.3, namely the
production of {\it new} unreliability within a module as measured by the
positivity of fiber Lyapunov exponents.  Mathematically, site
reliability (or unreliability) is reflected in the marginals of
$\mu_\omega$ at the site in question, while the unreliability produced
within a module is reflected in the dynamics and conditional measures of
$\mu_\omega$ on fibers.  Naturally, when a module produces unreliability
in the sense of Sect. 3.3, its sites will also show unreliability in the
sense of Sect. 4.1, and one cannot separate what is newly produced from
what is passed down from upstream.

%%%%%%%%%%%%%%%%%%%%%%%%%%%%%%%
%%%%%%%%%%%%%%%%%%%%%%%%%%%%%

\section{Examples:  reliability and unreliability in small networks}
\label{Examples of small networks}

In \paperone, we carried out a detailed study of a coupled oscillator pair.
One may seek a similar understanding of other small networks,
but any classification grows
rapidly in complexity with the number of sites. In this section, we do
not attempt a systematic study, but provide a few examples of small
networks exhibiting various behaviors.

In all the examples in this section, $\omega_i \in [0.95, 1.05]$.
%\note{Added}

\subsection{The $N$-ring}
Our first class of examples is a direct generalization of the 2-loop
studied in \paperone, namely networks of ``ring'' type:
\begin{equation}
\label{n-ring}
\includegraphics[scale=0.6]{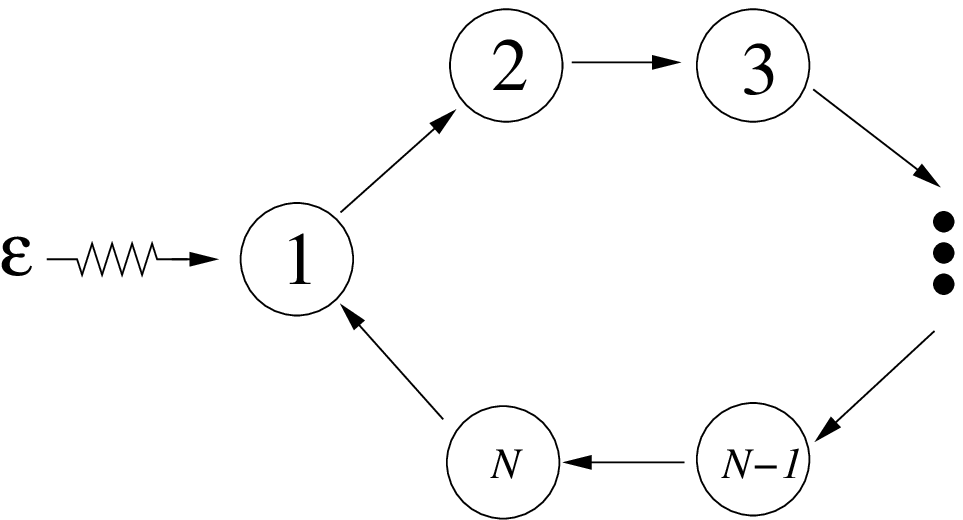}
\end{equation}
The main point of this example is to demonstrate that the system above
exhibits both reliable and unreliable dynamics.
Along the way, we also briefly illustrate two interesting phenomena
about which we have little explanation.

\bigskip
\noindent {\bf Example 1 }{\it (Amplification effects along loop?)} For
$N=2,3, \cdots, 7$, we fix $a_{i,i+1}$ for $i=1,\cdots, N-1$, and let
$a_{N,1}$ vary. Our findings can be summarized as follows:

\begin{itemize}
\item[(a)] With all the $a_{i,i+1} \approx 1$, the system is {\it reliable}
for $a_{N,1} \in [0,2]$ and $\eps<2$.
\item[(b)] For $\eps=1$ and $a_{i,i+1} \sim \frac{1}{N}$ suitably chosen, the system
is {\it unreliable} for $a_{N,1} \in [0.6,2]$.
\end{itemize}

%%%%%%%%%%%%%%%%%%%%%%%%%%%%%%%%
\begin{figure}
% src: ~/hopf/rely1.data/study07/2007-07-12=with-forcing.d/paper-figs.ss
\begin{center}
\begin{tabular}{cc}
\includegraphics[bb=0in 0in 2.5in 2in]{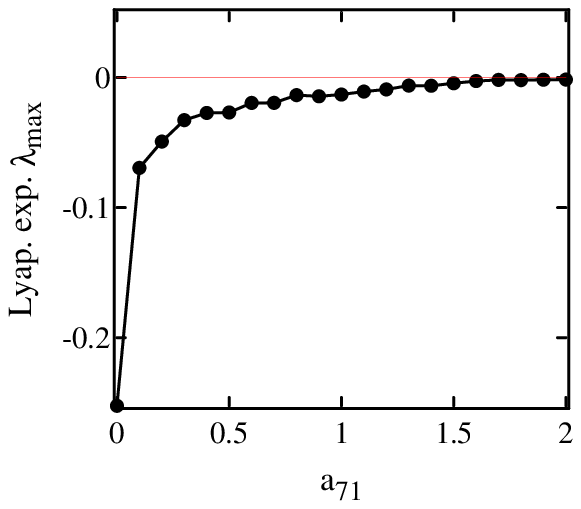}&
\includegraphics[bb=0in 0in 2.5in 2in]{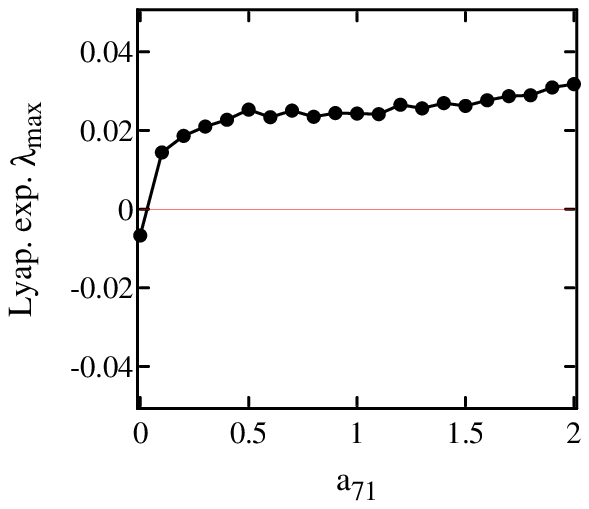}\\\\
(a) With $a_{i,i+1}\approx 1$&(b) With $a_{i,i+1}\sim 1/N$\\\\
\end{tabular}
\end{center}
\caption{Amplification effect in $N$-ring, with $N=7$.  We show the
Lyapunov
  exponent $\lmax$ as a function of $a_{N,1}$. In (a), the coupling
  constants $a_{i,i+1}$ for $i=1,2,\cdots,N-1$ are all $\approx 1$.  In
  (b), they are $\sim 1/N$ (see text).  In both plots, $\eps=1$.}
\label{n-ring scaling}
\end{figure}
%%%%%%%%%%%%%%%%%%%%%%%%%%%%%%%%

\noindent Results of numerical simulations for $N=7$ are shown in
Fig.~\ref{n-ring scaling}. For case (a) we randomly choose
 $\omega_j$ and $a_{i,i+1} \in [0.95, 1.05]$ for $1 \le j \le N$ and
$1 \le i <N$. For case (b), we start the same way, but multiply all the
$a_{i,i+1}$ chosen by a constant $c$. As we tune $c$ from $1$ down to
$0$, we find that in each case, pictures similar to that in
Fig.~\ref{n-ring scaling}(b) are produced for a range of $c \sim
\frac{1}{N-1}$ where $N$ is the size of the ring. Simulations were
repeated for different choices of $\omega_j$ and $ \tilde a_{i,i+1}$,
to similar effect. The exact $c$'s that produce the most unreliable
behaviors, however, vary with the choices of $\omega_j$ and
$a_{i,i+1}$; $\frac{1}{N-1}$ is only a rough approximation.

Details aside, the following trend as
$N$ increases from $2$ to $7$ is unmistakable:
Given that we use roughly equal ${a}_{i,i+1}$ for $1 \le i <N$
and set $a_{N,1} \approx 1$, the values of $a_{i,i+1}$ for which
unreliable dynamics occur decrease steadily as $N$ increases.
It is as though coupling effects are amplified along the loop,
giving rise to an effective coupling $``a_{1,N}" \sim 1$.
(For $N=2$, the system is known to be most unreliable when
$a_{12}$ and $a_{21}$
are roughly comparable; see \paperone.)
Whatever the explanation, this $N$-ring example
promises to be interesting (but we do not pursue it further here).

%%%%%%%%%%%%%%%%%%%%%%%%%%%%%%%%%
%\begin{figure}
%% src: ~/hopf/rely1.data/study07/2007-07-12=with-forcing.d/paper-figs.ss
%\begin{center}
%\includegraphics[bb=0in 0in 2.5in 2in]{graphics/four-ring-production-unscaled.eps}\qquad%
%\includegraphics[bb=0in 0in 2.5in 2in]{graphics/four-ring-cancel.eps}
%\end{center}
%
%\caption{Cancellation effect in 4-ring.  Both panels show the Lyapunov
%  exponent $\lmax$ as a function of $a_{41}$.  {\bf Left:} all other
%  coupling constants are $a_{i,i+1}\approx 1$.  {\bf Right:} we set
%  $a_{12}\approx a_{34}\approx 1$ and $a_{23} \approx -1$.  (Exact
%  values are $a_{12}=1.02, a_{23}=-0.98, a_{34}=1.05$.)  In all figures,
%  we set $\eps=1$.}
%\label{4-ring cancellation}
%\end{figure}
%%%%%%%%%%%%%%%%%%%%%%%%%%%%%%%%%

\bigskip
\noindent {\bf Example 2 }{\it (Cancellation effect by couplings of
  opposite signs?)} \noindent Given the conjectured amplification effect
just described, it is natural to ask whether negative coupling terms
along the loop would diminish the effects of positive coupling, and {\it
  vice versa}. This idea is validated in simulations.  For example, for
the 4-ring we find (as asserted above) that $\lmax<0$ when the coupling
constants are all $\approx 1$; the plot is qualitatively similar to
Fig.~\ref{n-ring scaling}(a). However, if we flip the sign of any one of
the 4 couplings to a value of $\approx -1$ while leaving the rest at
$\approx 1$, we then obtain $\lmax>0$ and plots qualitatively similar to
Fig.~\ref{n-ring scaling}(b).
%Notice
%that this cancelation idea is consistent with the amplification
%conjectured above (compare with Example 1(a)).

%%%%%%%%%%%%%%%%%%
\subsection{2-loops driving 2-loops}
\label{s.2loop}

Here we study a new situation, in which one unreliable module drives
another.  Consider the system
\begin{equation}
\includegraphics[scale=0.6]{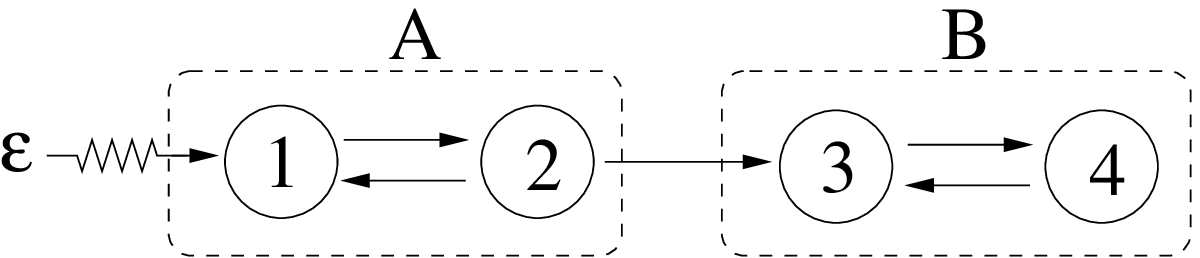}
\label{twobytwo}
\end{equation}
We think of this network as consisting of two modules: the (1,2)-subsystem
as Module A and the (3,4)-subsystem as Module B. From
\paperone, we have a great deal of information about these
2-oscillator systems individually.
Here, we investigate the unreliability produced within Module B
when driven by Module A.  Representing system \eqref{twobytwo} as a skew product with
2-dimensional fibers over a 2-dimensional base (as in
Sect.~\ref{sect2.3}), we let $\lmax^{fib}$ denote the largest fiber
Lyapunov exponent for Module B.

\begin{figure}[t]
\begin{center}
% src: ~/hopf/rely1.data/study06/2007-07-02b.d
\begin{tabular}{cc}
\includegraphicsbw{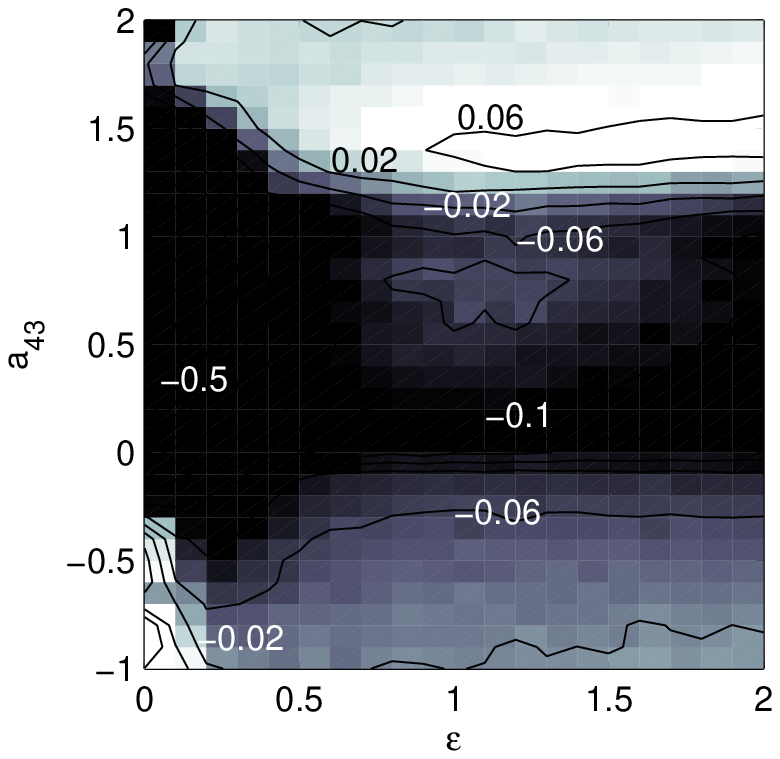}{bb=0in 0in 3.4in 3.2in}&
\hspace*{-12pt}\includegraphicsbw{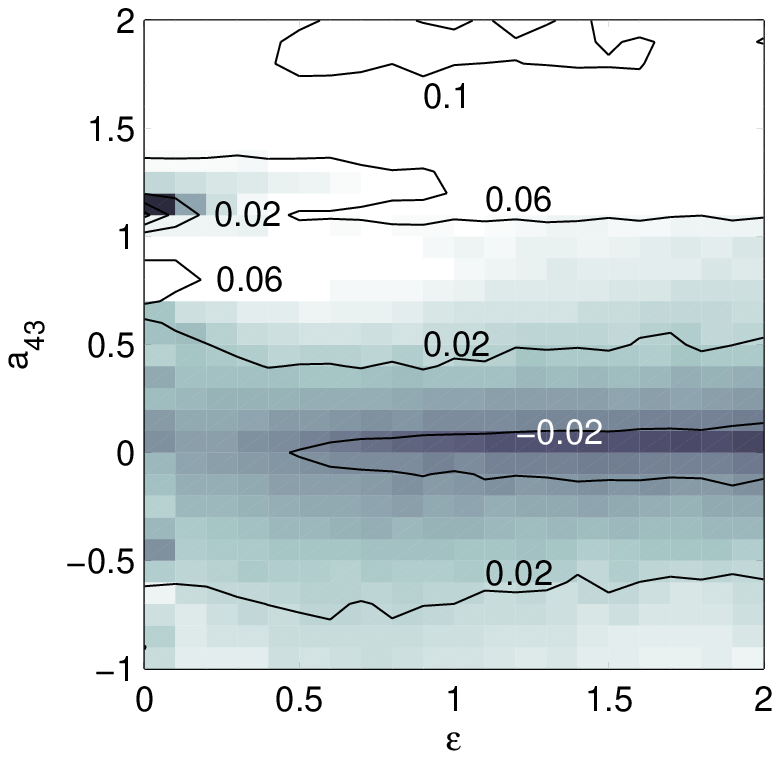}{bb=0in 0in 3.4in 3.2in}%
\hspace*{-18pt}\includegraphicsbw{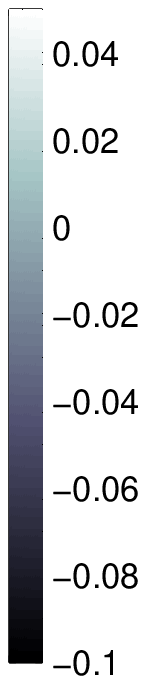}{bb=0in 0in 0.6in 3.2in}\\
(a)& (b)\\
\end{tabular}
\end{center}
\caption{The maximum fiber exponent $\lmax^{fib}$ for Module B in system
  \eqref{twobytwo}, as a function of $\eps$ and $a_{43}$.  In (a), the
  parameters are $a_{12}=1.1$, $a_{21}=0.8$, $a_{23}=0.5$, and
  $a_{34}=1$.  In (b), the parameters are $a_{12}=-1.1$, $a_{21}=-1.2$,
  $a_{23}=0.95$, and $a_{34}=1$.  In both cases, we use $\omega_1=1.03$,
  $\omega_2=0.98$, $\omega_3=1$, and $\omega_4=1.1$.  {\bf Remark on
    plots:} We have chosen the dynamic range in shading the figures to
  allow meaningful comparison of figures; a side effect is that some
  contour lines may not be visible.  We always indicate the actual range
  of values through explicit labels. }
\label{model c plots}
\end{figure}

We fix two sets of parameters for which Module A is known to be
unreliable.  For one set, Module A has mutually ``excitatory''
coupling, with $a_{12},a_{21}>0$; for the other, Module A
has mutually inhibitory coupling, with $a_{12},a_{21}<0$. We then
perform numerical experiments (see Fig.~\ref{model c plots}) in which
we fix $a_{34}=1$ and vary $a_{43}$ and the system input amplitude
$\eps$.  Our findings are:
%We find that $\lmax^{fib}$ is affected nontrivially by whether Module A
%consists of a

\begin{itemize}

\item[({\it i})] For both choices of  parameters for Module A, unreliability
is sometimes produced within Module B and sometimes not,
depending on $a_{43}$.

\item[({\it ii})] For most values of $a_{34}, a_{43}$ and $\eps$,
Module B produces more unreliability ({\it i.e.}, $\lmax^{fib}$ is larger)
%\note{rephrased -- sounds funny to say a module is mutually inhibitory}
when the oscillators in
Module A are mutually inhibitory rather than excitatory.
\end{itemize}

\noindent Both phenomena appear to be robust.
We do not have an explanation for ({\it ii}), and will revisit ({\it i})
in Sect.~6.

The situation at $\eps=0$ is shown in the leftmost column of each of
the two panels in Fig.~\ref{model c plots}. As with the 3-ring in
Sect.~2.2, $\lmax^{fib}$ at $\eps=0$ varies sensitively with parameter.
The production of intrinsic network chaos within Module B is evident
for some parameters in both cases. For some parameter settings, this
chaos is suppressed; see the bottom
left panel.  %\note{deleted ``suff lg" $\eps$: small actually}
%The suppression of this chaos is evident in the bottom of the
%left panel.
Overall, notice the strong influence of the $\eps=0$ dynamics of Module
B on its tendency to produce unreliability when the external stimulus
is turned on.

%\bigskip
%\noindent {\bf Remarks.} The unreliability produced within a module in
%a network, as discussed above, is related to the ideas in
%Sect.~\ref{site
%  reliability} as follows: the concern in Sect.~\ref{site reliability}
%is whether or not unreliability created upstream ({\it e.g.} in Module
%A) propagates, {\em i.e.} can be observed downstream, whereas
%$\lmax^{fib}>0$ implies that {\it new} unreliability is created.
%Mathematically, site entropy is reflected in the marginals of
%$\mu_\omega$ at the site in question, while the unreliability produced
%by Module B is reflected in the conditional measures of $\mu_\omega$ on
%fibers.  Of course, if $\lmax^{fib} > 0$ and we find that, say, site 4
%is is unreliable (in the sense of having a uniform site distribution,
%as in Sect.~\ref{site reliability}), it would not be possible to
%separate the unreliability propagating from Module A from the
%unreliability produced within Module B by looking at the site-4
%distribution alone.

%%%%%%%%%%%%%%%%%%%%%%%%%%%%%%
%%%%%%%%%%%%%%%%%%%%%%%%%%%%%%
\section{Modules in isolation versus as part of larger system}
\label{isolation}

Now that we have discussed both propagation of unreliability between
modules and generation of unreliability within modules, we return to
the strategy suggested in Sect.~\ref{sect2.3}: to study the reliability
of networks by analyzing their component modules separately.  An
obvious benefit of this strategy is that smaller modules are easier to
test. However, the strategy will only be successful if the reliability
of modules in isolation gives a good indication of their reliability
when they are embedded in larger systems.\footnote{This type of issue
arises
  often in coupled nonlinear systems; it is very important yet seldom
  understood.}

Pictorially, the problem we face can be represented as follows:

\vspace{3ex}
\begin{equation}
\includegraphics[scale=0.5]{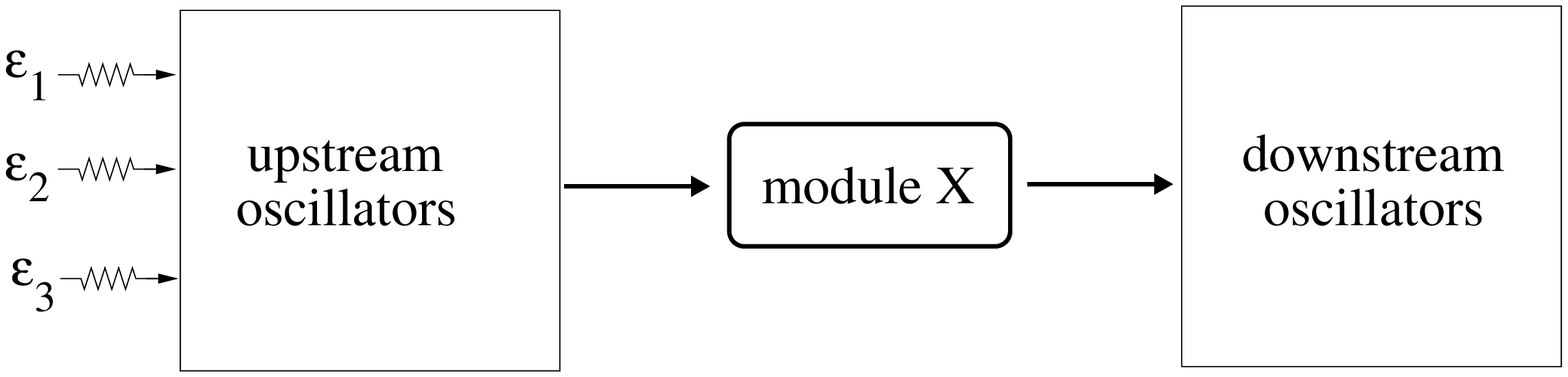}
\vspace{3ex}
\label{f.blackbox}
\end{equation}
Suppose our network can be decomposed into three parts: Module X, which
is our module of interest, a (possibly large) component upstream from
Module X, and a (possibly large) component downstream from Module X.
The question is: {\it Can the reliability properties of Module X as a
  subsystem of this larger network be predicted by its reliability
  properties in isolation?}

First we need to clarify what we mean by the ``reliability of a module
in isolation''. We refer here to the reliability of the module (without
the rest of the network) receiving stimuli with a simple, generic
statistical structure; in this paper, this statistical structure is
taken to be white noise. To simulate the correct phase-space
geometry, it is important that the white noise be injected in a way
that mimics network
conditions. For example, if oscillators 1, 2, and 4 within Module X
receive input from the rest of the network, then in analyzing the
reliability properties of Module X in isolation, we must provide
stimuli to the corresponding oscillators.\footnote{To illustrate this
point, consider a 2-oscillator system with unequal feedback/feedforward
couplings: providing a stimulus to oscillator 1 leads to horizontal
perturbations in $(\tht_1,\tht_2)$-space, whereas
stimulating oscillator 2 leads to vertical perturbations. These two types of perturbations have different relations to the geometry of the intrinsic
dynamics and can have different effects.} One should also use comparable
forcing amplitudes, although it is less clear what exactly that means.

At the heart of this question is the following issue: A module embedded
in a network may receive input from oscillators upstream in the form of
coupling impulses; it may also hear directly from the external stimuli.
Inputs from other oscillators resemble a point process of impulses with
statistics somewhere between Poisson and periodic; moreover, when an
oscillator receives kicks from multiple sources, these kicks may be
correlated to varying degrees.  The question is then whether
reliability properties of a module are sufficiently similar under these
different classes of inputs that they can be predicted from studies
using white noise stimuli.

Let us return now to the situation depicted in the diagram above.
For a quantitative study, we first compute $\lmax$ for Module X in isolation, subjected to white noise in a suitable range of amplitudes.
These numbers will be compared to $\lmax^{fib}$ defined as follows:
Consider the skew product in which the dynamics of the
oscillators upstream from Module X are represented in the base
and the dynamics within Module X are represented on the fibers.
(The oscillators downstream from Module X are irrelevant.)
The number $\lmax^{fib}$ is the largest fiber Lyapunov exponent
(see Sect.~\ref{sect2.3}).
The following abbreviated language
will be used in the remainder of this paper: when
we speak of {\em the unreliability produced by a module}, we refer
to the unreliability produced within that module  when it operates
as a subsystem in the larger network.  The same abbreviation applies
to intrinsic network chaos in a module.

\bigskip
\noindent {\bf Proposal:} {\it Whether unreliability is produced within
  Module X, {\em i.e.} whether $\lmax^{fib} > 0$ when $\eps > 0$, is
  strongly influenced by
\begin{itemize}
\item the reliability properties of Module X in isolation, and
\item the intrinsic dynamics within Module X when embedded in the full network, {\em i.e.}, $\lmax^{fib}$ at $\eps=0$
\end{itemize}
Part (b) is especially relevant at small $\eps$.}

\bigskip The proposal above is intended to give only a rough indication
of the hills and valleys of the reliability profile of the module in
question; no quantitative predictions are suggested.  Part (b) of the
proposal is largely a statement of continuity: when subjected to weak
external stimuli, intrinsic network properties come through as
expected. The rationale for (a) is that we have seen in a number of
situations in general dynamical systems theory that the response of a
system to external forcing depends considerably more strongly on its
underlying geometry than on the type of forcing, especially when there
are elements of randomness in the forcing, {\em e.g.} Poisson kicks and
white-noise forcing have qualitatively similar effects (though purely
periodic forcing can do peculiar things).  A systematic study comparing different types of forcing is carried out
in \cite{ly07}.

%%%%%%%%%%%%%%%%%%%%%%%%%%%%%%%%
\begin{figure}
\begin{center}
\includegraphicsbw{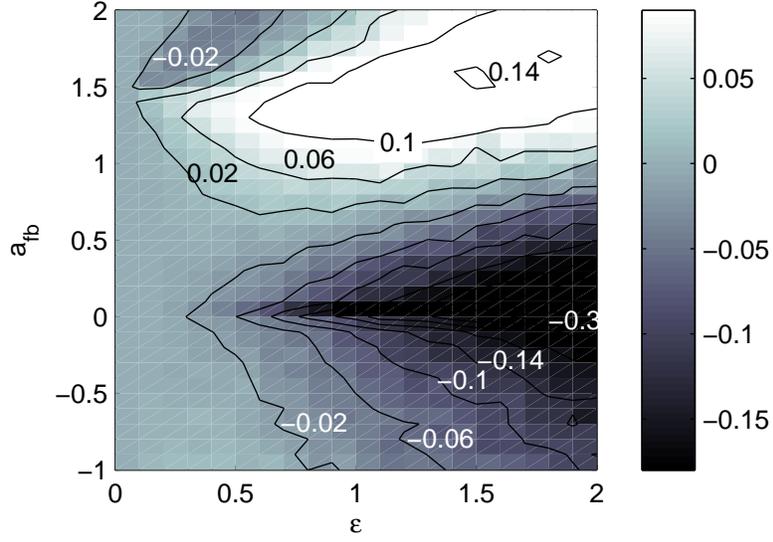}{bb=0in 0in 4.25in 3in}\\
(a) Two oscillators in isolation\\
% src: ~/hopf/rely1.data/study06/2007-07-02b.d
\begin{tabular}{p{2.8in}p{2.8in}p{0.4in}}
\includegraphicsbw{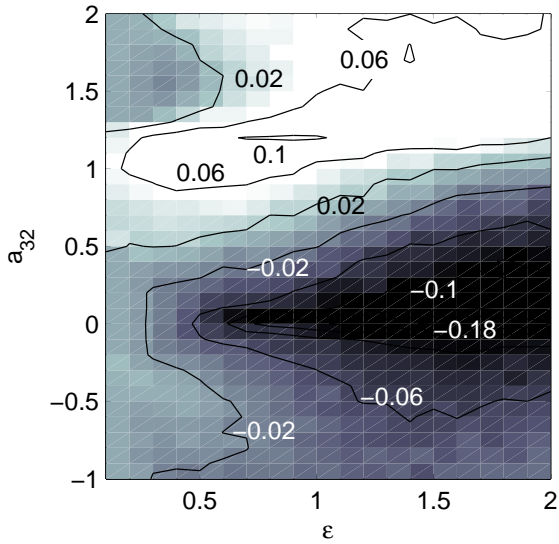}{bb=0in 0in 3.4in 3.2in,scale=0.94}&
\includegraphicsbw{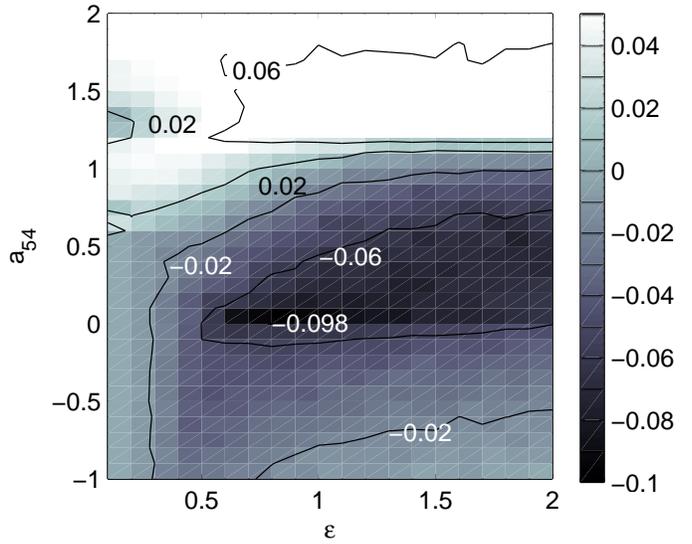}{bb=0in 0in 3.4in 3.2in,scale=0.94}&
\includegraphicsbw{graphics/model_abc_colorbar}{bb=0in 0in 0.6in 3.2in,scale=0.94}\\
\centering{(b) Network A}&\centering{(c) Network B}&\\
\end{tabular}
\end{center}
\caption{Reliability of Module X in isolation and as an embedded subsystem.
  In (a), we show $\lmax$ for Module X in isolation.  The parameters are
  $\omega_1=1$, $\omega_2=1.1$, and $\aff=1$. The corresponding modules
  in Networks A and B are given the same parameters.  The other parameters
  are as follows: for Network A, we use $\omega_1=0.97$, $a_{12}=0.7$.
  For Network B, we draw $\omega_i\in[0.95,1.05]$ and set $a_{i,4}=
  \frac{1}{3}\cdot(0.6, 0.8, 0.9)$ and $\eps_1=\eps_2=\eps_3=\eps$.
       }
\label{Models A and B}
\end{figure}
%%%%%%%%%%%%%%%%%%%%%%%%%%%%%%%%

\medskip
To support our proposal, we present below the results of a study in
which the 2-cell system studied in \paperone, {\it i.e.},

\ \\
\begin{center}
\vspace{-0.5in}
\includegraphics*[bb=64 0 194 39,scale=0.6]{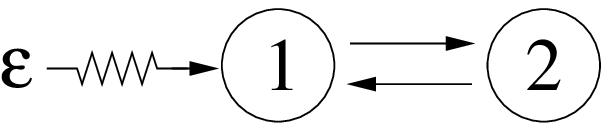}
\end{center}
\vspace{0.1in}

\noindent is used as the embedded module, that is, as Module X in the
framework above.  The three networks
used in this study, with Module X enclosed in a dotted line, are:
\vspace{0.1in}

Network A: \\
\begin{center}
\vspace{-0.5in}
\includegraphics[scale=0.6]{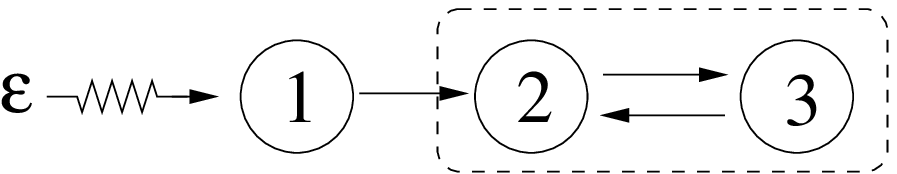}
\end{center}
\vspace{0.1in}

Network B:\\
\begin{center}
\vspace{-0.5in}
\includegraphics[scale=0.6]{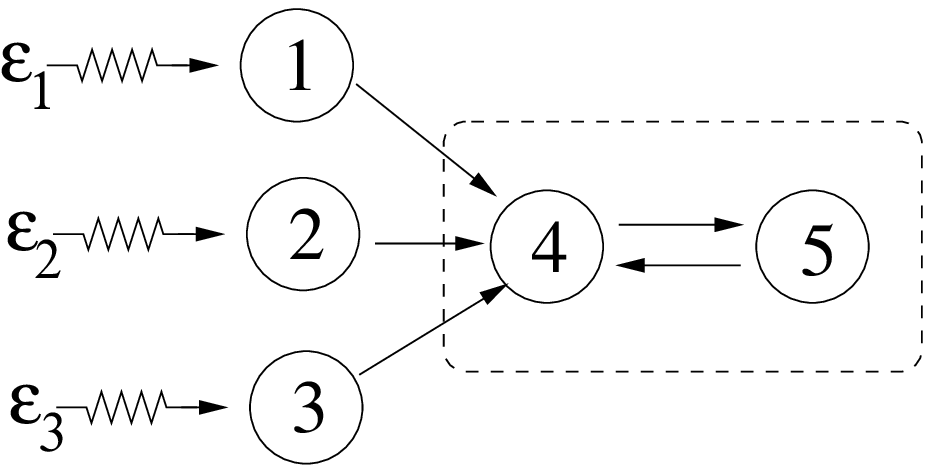}
\end{center}
\vspace{0.1in}

Network C:\\
\begin{center}
\vspace{-0.5in}
\includegraphics[scale=0.6]{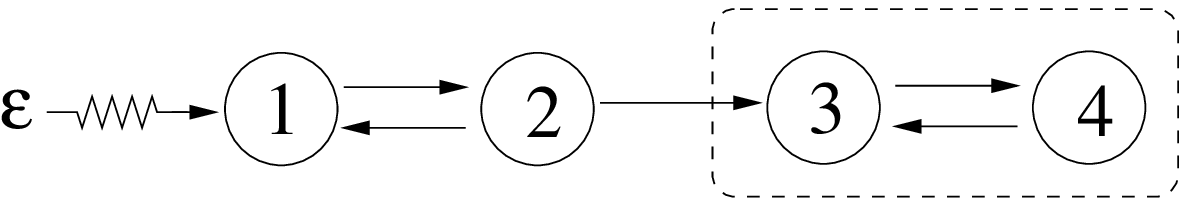}
\end{center}
\vspace{0.1in}

\noindent In Networks A and B, the oscillators that feed into Module X
are themselves reliable. Network C is the system described in
Sect.~\ref{s.2loop}; as described there, parameters are chosen so that
the subsystem (1,2) is unreliable. In each case, the feedforward
coupling within Module X is set equal to 1, and the largest
 fiber exponent is computed as a
function of the feedback coupling and $\eps$. Sample plots for Networks A
and B are shown in Fig.~\ref{Models A and B}. For comparison, we have
also included the plot for the 2-oscillator system in isolation
(Fig.~12 in \paperone), {\it i.e.} receiving a white noise input of
strength $\eps$ at oscillator 1, to mimic the configuration in Networks
A-C.  Plots for Network C are in Fig.~\ref{model c plots}.

The resemblance of Figs.~\ref{Models A and B}(b) and (c) to
Fig.~\ref{Models A and B}(a) is quite striking given the rather
different inputs received by the Module X in isolation {\it vs.} embedded in
Network B. This is evidence in favor of part (a) of the
proposal above. Network C also gives qualitatively similar reliability
profiles, although overall it is considerably more reliable for one set
of parameters governing oscillators (1,2) than the other:  see
Figs.~5(a) and (b).  As suggested by the second part of the proposal,
there is a strong correlation between all of these overall tendencies
and network dynamics at $\eps=0$, sensitive as the latter may be.

Here as in many other simulations not shown, we have found that even though
the reliable and unreliable regions may shift and the magnitudes of
$\lmax^{fib}$ may vary, the general tendency is that the reliability of
Module X continues to resemble its reliability in isolation, that is,
when it receives white noise inputs. We take these results to be
limited confirmation of our proposal.  In sum:

\bigskip
\noindent {\bf Observation 4:} {\em The reliability profile of the
  two-oscillator system in isolation found
  in \paperone\ is a reasonable guide to its reliability properties as a
  module embedded in a larger network.}

\medskip
\noindent
It remains to be seen whether this holds for other modules.

%%%%%%%%%%%%%%%%%%%%%%%%%%%
%%%%%%%%%%%%%%%%%%%%%%%%%%
%\section{Network reliability via modular decomposition:
%an example}
%\label{lambda}

\section{An illustrative example}
%\addcontentsline{toc}{section}{Conclusions}

We finish with the following example, which illustrates many of
the points discussed in this paper.

\begin{figure}[t]
\begin{tabular}{p{3in}p{1in}p{1.3in}}
\includegraphics[scale=0.85]{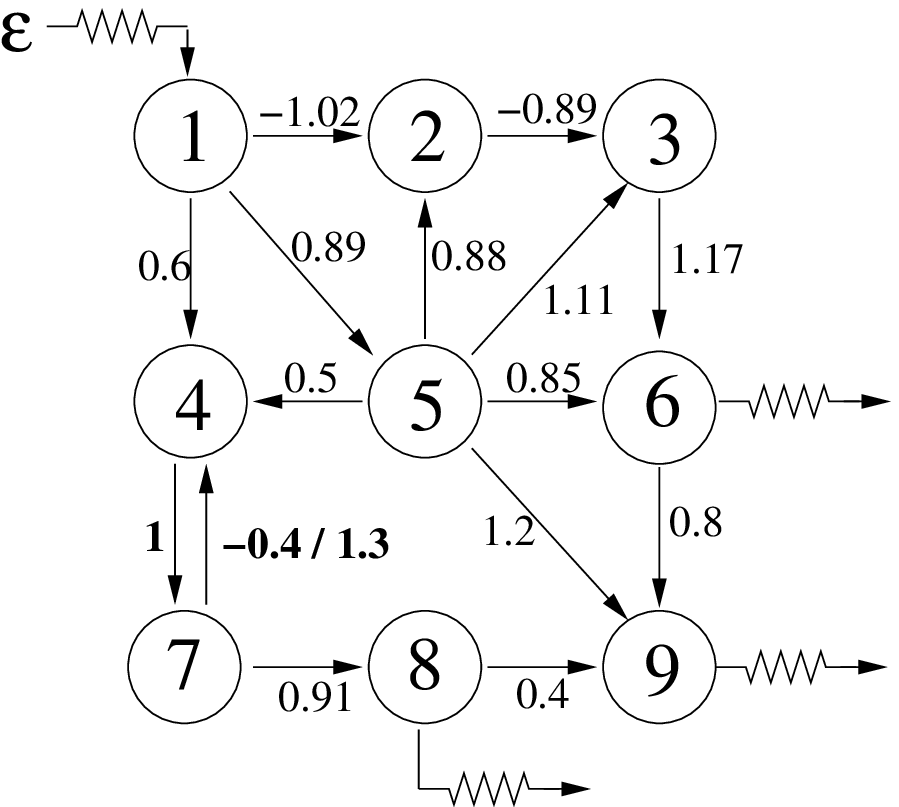} &&
\includegraphics[bb=72 -72 257 164,scale=0.7]{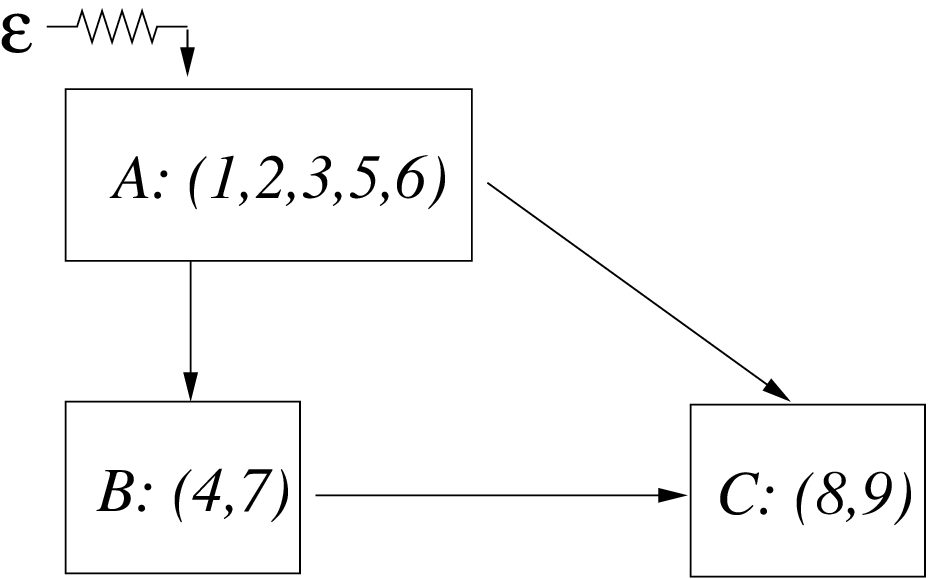}\\
\centering{(a) Full network} &&
\centering{(b) Quotient graph}\\
\end{tabular}\vspace{12pt}
\caption{Example of a larger network and its quotient graph.  In (a), we
  have labelled the edges with a sample of coupling constants.  The
  $\omega_i$ are drawn from $[0.95,1.05]$.}
\label{complicated example}
\end{figure}

The network depicted in Fig.~\ref{complicated example}(a) is made up of
9 oscillators. It receives a single input stimulus through oscillator 1
and has ``output terminals'' at sites 6, 8 and 9 -- it is here that we
will assess the response of the network. We first discuss what to
expect based on the ideas above. This is then compared to the results
of numerical simulations.

A cursory inspection tells us that this network is acyclic except for
the subsystem (4,7). The finest decomposition
that yields an acyclic quotient, then, is to regard each oscillator as a module
except for (4,7), which must be grouped together as one. We choose, however,
to work with a coarser decomposition, in which the (1,2,3,5,6)-subsystem
is viewed as Module A,
the (4,7)-subsystem as Module B, and the (8,9)-subsystem as Module C.
Identifying the sites within each of these modules produces an acyclic
quotient, as shown in Fig.~\ref{complicated example}(b).

Since Module B is the only module that is not itself acyclic and hence
 that is capable of generating
unreliability, our results from Sect.~\ref{acyclic} tell us that
$\lmax$ for the entire network is $\le 0$ if and only if no
unreliability is produced in Module B. In fact, since there are no
freely rotating oscillators in the system, we expect $\lmax$ to be $<0$
if Module B behaves reliably. The behavior of Module B hinges a great
deal on (i) the couplings $a_{47}$ and $a_{74}$, which determine its
reliability in isolation, and (ii) intrinsic network properties in the
two Modules A and B together, esepcially when $\eps$ is small; see
Sects.~\ref{intrinsic chaos} and~\ref{isolation}. With regard to (i),
the reliability profile of the 2-oscillator system complied in
\paperone\ is handy. If unreliability is produced in Module B, then we
expect to find sites 8 and 9 to be unreliable, with a lower reading of
site entropy at site 9 than 8 due to the stabilizing effects of Module
A; see Sect.~4. This is what general theory would lead one to predict.

We now present the results of simulations.

First we confirm that Module A alone is reliable as
predicted:\footnote{All Lyapunov exponents presented in this section
  have standard errors of $\leq 0.004$ as estimated by the method of
  batched means.  By the Central Limit Theorem, this means the actual
  $\lmax$ should lie within $\approx 2.5\times 0.004 = 0.01$ of the
  computed value with $\gtrsim 99\%$ probability.}  In addition to the
$a_{ij}$ given in Fig.~\ref{complicated example},
%\note{added phrase}
we randomly drew coupling constants from $\pm[0.8,1.2]$ and frequencies
$\omega_i\in[0.95,1.05]$, and find that $\lmax$ for Module A ranges from
roughly $-0.3$ to $-0.07$ when $\eps=1$. Site distributions for the
sites in Module A are, as predicted, well-localized.  For a majority of parameters tested, 90\% of an ensemble of
  $10^4$ uniformly-chosen initial conditions has collapsed into
  a cube of side length $\leq 10^{-2}$ after t=60-110; in all our simulations,
  the ensemble collapses into a cube of side length $\lesssim
  10^{-7}$ after $t\approx 1000$.  In
particular, the ``output terminal'' at site 6 is always reliable.

Turning now to the reliability of Module B, we fixed the coupling constants
as shown in Fig.~\ref{complicated example}, with $\omega_i\in[0.95,1.05]$
and $\eps=1$,
% \begin{align*}
% &\omega_1 = 0.957, \omega_2 = 0.989, \omega_3 = 1.043, \omega_5 =
% 0.970,
% \omega_6 = 0.982,\\
% &a_{12} = -1.023, a_{23} = -0.893, a_{15} = 0.885, a_{52} = 0.883,
% a_{53} = 1.110, a_{36} = 1.170, a_{56} = 0.847.
% \end{align*}
and ran simulations for the following two sets of parameters:
\begin{itemize}
\item $a_{47}=1, a_{74}=-0.4$: this case is predicted to be reliable
\item $a_{47}=1, a_{74}=1.3$: this case is predicted to be unreliable
\end{itemize}
The predictions above are based on the behavior of the two-oscillator
network ``in isolation'' (receiving white-noise stimulus of amplitude 1, see \paperone), and on the results of Sect.~6.
The following table summarizes the reliability
properties of Module B, both in isolation and when embedded within the
network:

\begin{center}
\begin{tabular}{lp{0.8in}||c||c|c}

\multicolumn{5}{c}{\bf Module B Response}\\\\

&&&\multicolumn{2}{c}{\em Embedded in network}\\

  && \raisebox{1.5ex}[0pt][0pt]{\em In isolation} & $\eps=0$ & $\eps=1$\\\hline

\raisebox{-1.5ex}[0pt][0pt]{(a)}& $a_{47}=1$,\newline $a_{74}=-0.4$ &
\raisebox{-1.5ex}[0pt][0pt]{$\lmax=-0.07$} &
\raisebox{-1.5ex}[0pt][0pt]{$\lmax^{fib}=0.014$} &
\raisebox{-1.5ex}[0pt][0pt]{$\lmax^{fib}=-0.15$}\\\hline

\raisebox{-1.5ex}[0pt][0pt]{(b)}& $a_{47}=1$,\newline $a_{74}=1.3$&
\raisebox{-1.5ex}[0pt][0pt]{$\lmax=0.13$} &
\raisebox{-1.5ex}[0pt][0pt]{$\lmax^{fib}=0.076$}  &
\raisebox{-1.5ex}[0pt][0pt]{$\lmax^{fib}=0.097$} \\

\end{tabular}
\end{center}

\noindent
Note that these values are consistent with the
Proposal in Sect.~\ref{isolation}: the behavior of the module at
$\eps=1$ when embedded within the network is effectively determined by
its behavior in isolation.  Furthermore, by Prop.~\ref{c.decomp}, we
know the Lyapunov exponent $\lmax$ of the {\em entire network} is equal
to $\lmax^{fib}$ of Module B in case (b), and is $\geq\lmax^{fib}$ in
case (a).

Finally, we study site distributions at sites 8 and 9.
In case (a), we find again that computed site distributions are well
localized.  In case (b), we find the expected
evidence of propagated unreliablility and interference, with
${H}(8) = -0.3$ and ${H}(9) = -0.7$.

\medskip
In summary, the results of our simulations are entirely consistent with
predictions based on the ideas developed in this paper.

\section*{Conclusions}
\addcontentsline{toc}{section}{Conclusions}

Understanding and predicting the reliability of responses for a general
network is a daunting task.  In this paper, we have limited ourselves to
the simplest intrinsic dynamics, namely those of phase oscillators with
pulsatile coupling. In this context, we believe we have raised and
addressed a few basic issues for an interesting class of networks. Our main findings are:

\medskip
\begin{enumerate}

\item {\it Acyclic networks}, {\it i.e.},
networks with a well defined direction of information flow, {\it are

    \begin{itemize}
    \item never chaotic in the absence of inputs, and
    \item never unreliable when external stimuli are presented.
    \end{itemize}
}
\smallskip
\item {\it The mathematical analysis of the acyclic case can be extended to treat
networks with a modular decomposition and acyclic quotient, \it i.e.},
networks for which the nodes can be grouped into modules, with
inter-module connections being acyclic. {\it These networks are fully
capable of reliable and unreliable dynamics.}
% and the simplest such
%module is the two- oscillator system studied in Paper I of this series.

\medskip
\item {\it For networks with a nontrivial module decomposition, production of unreliability can be
localized to specific modules, and is indicated by positive fiber
Lyapunov
exponents.}% in our skew-product representation of network dynamics.}

\medskip
\item {\it Once it is produced, unreliability will propagate: it can be attenuated
but not completely removed; without intervention
it actually grows for sites farther downstream.}

\medskip
\item {\it There is evidence that the reliability of a module
in isolation may give a good indication of its behavior when embedded
in a larger network.} If true, this would simplify
the analysis of networks that decompose into relatively simple modules
with acyclic connections.
\end{enumerate}

\medskip \noindent
A general understanding of the issue in item 5  is not yet in sight,
but our limited testing -- using the 2-oscillator module studied in the
companion paper \paperone \ -- produced encouraging results.

We hope that the findings above will be of some use in
concrete applications in neuroscience and other fields.

\bigskip \textbf{Acknowledgements:} K.L. and E.S-B. hold NSF Math. Sci.
Postdoctoral Fellowships and E.S-B. a Burroughs-Wellcome Fund Career
Award at the Scientific Interface; L-S.Y. is supported by a grant from
the NSF.

\bibliographystyle{plain}
\bibliography{main}

\end{document}